\documentclass[sigconf]{acmart}
\pdfoutput=1

\makeatletter
\let\@authorsaddresses\@empty
\makeatother

\usepackage{tikz}
\usepackage{epsfig,endnotes}
\usepackage{type1cm}	
\usepackage{graphicx}     
\usepackage{xspace}     
\usepackage{booktabs}     
\usepackage[tableposition=top,font={scriptsize,bf}, skip=7pt]{caption}     
\usepackage{siunitx}	
\usepackage{framed}
\usepackage{newfloat}
\usepackage{amsmath, eurosym}
\usepackage{enumitem}
\usepackage[vlined,linesnumbered,ruled,noend]{algorithm2e}
\usepackage{lscape}
\usepackage{array}
\usepackage{pifont}
\usepackage{color, colortbl}
\usepackage{hyperref}
\usepackage{multirow}
\usepackage{url}

\definecolor{heraldBlue}{rgb}{0.0,0.0,0.8}

\def\ttfntsize{9}
\makeatletter
\let\oldtexttt\texttt
\let\texttt\@undefined
\makeatother
\newcommand{\texttt}[1]{\fontsize{\ttfntsize}{\ttfntsize}\oldtexttt{#1}}
\makeatletter
\let\oldtt\tt
\let\tt\@undefined
\makeatother
\newcommand{\tt}{\fontsize{\ttfntsize}{\ttfntsize}\oldtt}

\newcommand{\etc}{etc.}
\newcommand{\eg}{e.g., }
\newcommand{\ie}{i.e., }
\newcommand{\etal}{et al. }

\newcommand{\toolname}{\emph{YourAdvalue}\xspace}
\newcommand{\rusers}{200\xspace}
\newcommand{\toolAvailable}{11 months\xspace}

\newcommand{\datasetyouradvalue}{12000\xspace}
\newcommand{\datasetawazza}{80k\xspace}

\newcommand{\point}[1]{\par\smallskip\noindent\textbf{#1} }

\begin{document}
\title{YourAdvalue: Measuring Advertising Price Dynamics without Bankrupting User Privacy}
\subtitle{Please cite our ACM Sigmetrics Publication, DOI: 10.1145/3491044}

 \author{Michalis Pachilakis}
 \affiliation{
 	\institution{University of Crete / FORTH}
 	\country{Greece}
 }

 \author{Panagiotis Papadopoulos}
 \affiliation{
 	\institution{Telefonica Research}
 	\country{Spain}
 }

 \author{Nikolaos Laoutaris}
 \affiliation{
 	\institution{IMDEA Networks Institute}
 	\country{Spain}
 }

 \author{Evangelos P. Markatos}
 \affiliation{
 	\institution{University of Crete / FORTH}
 	\country{Greece}
 }

 \author{Nicolas Kourtellis}
 \affiliation{
 	\institution{Telefonica Research}
 	\country{Spain}
 }

\renewcommand{\shortauthors}{Michalis Pachilakis et al.}

%
\copyrightyear{2021} 
\acmYear{2021} 
\setcopyright{acmlicensed}
\acmConference[ACM Sigmetrics'21]{Sigmetrics '21: ACM Meas. Anal. Comput. Syst}{December 2021}{XXXX}
\acmBooktitle{Proc. ACM Meas. Anal. Comput. Syst}
\acmPrice{15.00}
\acmDOI{10.1145/10.1145/3491044}
\acmISBN{978-1-4503-XXXX-X/18/06}



\begin{abstract}

The Real Time Bidding (RTB) protocol is by now more than a decade old.
During this time, a handful of measurement papers have looked at bidding strategies, personal information flow, and cost of display advertising through RTB.
In this paper, we present \toolname, a privacy-preserving tool for displaying to end-users in a simple and intuitive manner their advertising value as seen through RTB.
Using \toolname, we measure \textit{desktop} RTB prices in the wild, and compare them with desktop and \textit{mobile} RTB prices reported by past work.
We present how it estimates ad prices that are encrypted, and how it preserves user privacy while reporting results back to a data-server for analysis.
We deployed our system, disseminated its browser extension, and collected data from \rusers users, including \datasetyouradvalue ad impressions over \toolAvailable. 
 
By analyzing this dataset, we show that desktop RTB prices have grown $4.6\times$ over desktop RTB prices measured in 2013, and $3.8\times$ over mobile RTB prices measured in 2015.
We also study how user demographics associate with the intensity of RTB ecosystem tracking, leading to higher ad prices.
We find that exchanging data between advertisers and/or data brokers through \emph{cookie-syncronization} increases the median value of displayed ads by 19\%.
We also find that female and younger users are more targeted, suffering more tracking (via cookie synchronization) than male or elder users.
As a result of this targeting in our dataset, the advertising value (i) of women is $2.4\times$ higher than that of men, (ii) of 25-34 year-olds is $2.5\times$ higher than that of 35-44 year-olds, (iii) is most expensive on weekends and early mornings.

\end{abstract}

\maketitle

\section{Introduction}\label{sec:intro}

In the last few years, around 300 billions dollars are spent per annum in digital advertising globally~\cite{300billions}.
The dominant portion of this amount (84.9\% for the case of US) go to \emph{programmatic advertising}~\cite{programmatic2019}.
In programmatic advertising, real-time auctions that last less than 100 ms take place before a visitor finishes loading a web-page, to determine which advertisements they get to see.
The most popular medium for conducting such real-time auctions is the Real-Time Bidding (RTB) protocol, amounting for 90\% of all programmatically purchased ads~\cite{rtbperc} (the remaining percentage is split between Header-Bidding~\cite{header_bidding,cook2020inferring}, Programmatic Direct~\cite{progDirect} and Private Exchange buying (PMP)~\cite{privateEx}). 

The main promise of programmatic advertising and RTB is that visitors get to see only ads tailored to their actual interests, instead of generic ones that may be of little relevance. The key benefits of RTB include: (i) granular and targeted ad-buying, (ii) cost optimization, efficiency, speed, and (iii) enhanced transparency to both the advertiser and the publisher.
But \emph{what about the transparency to the factors that impacts the pricing dynamics and generates the data for these decision engines, \ie the user?}
Typically, users learn little, if any, regarding the prices that advertisers pay to buy the real estate of an ad-slot on their displays.

There are only two studies aiming to shed light on the pricing dynamics of the RTB ecosystem.
~\cite{lukasz2014selling-privacy-auction} reports cleartext RTB prices from artificial personas, as well as a handful of prices (400) from real users, to assess how ad prices vary across different DSPs in three countries.
~\cite{imcRTB2017} showed how to estimate the value of RTB winning bids, even when they are encrypted.
This was done by running purpose-built probe campaigns to obtain ground truth data about encrypted prices, that were then used to train a classifier that predicts encrypted prices based on features of users, ads, and domains.
To demonstrate the effectiveness of their classifier, the authors tested their approach on datasets from a proxy for mobile users.

In this paper, we present \toolname: a first-of-its kind transparency software~\cite{laoutaris2018data} for programmatic advertising over RTB.
\toolname allows end-users to learn in real-time while browsing, how much money the programmatic advertising ecosystem pays to capture their attention by rendering ads on their display. 
\toolname takes into account both cleartext and encrypted winning bids.
Users help in estimating the value of the latter by committing ad prices and auction metadata in a crowd-sourced database, in a privacy-preserving manner.
Data gets first anonymized, and then feature aggregation is applied, prior to their transmission towards the database.
This allows \toolname to re-train its classifiers, and improve its accuracy on estimating the most up-to-date encrypted prices.
By better understanding the value of their digital footprint, we expect that end-users will be able to make more informed choices about how they wish to participate in the emerging data economy. 
\toolname can also help data protection authorities, regulators, and watch-dog groups for signs of bias and discrimination against particular demographic groups~\cite{gonzalez2017fdvt, 10.1145/3366423.3380221}.

Over a period of \toolAvailable, we collected information from \datasetyouradvalue ads shown to \rusers users of \toolname.
This is, to the best of our knowledge, the largest desktop RTB data collected thus far ($30\times$ more impressions shown to real users than~\cite{lukasz2014selling-privacy-auction}).
We analyzed this dataset for factors affecting the prices (age, gender, geo-location, time, user tracking, etc.), studied the time-evolution of pricing with respect to past reports from~\cite{lukasz2014selling-privacy-auction,imcRTB2017}, and assessed the privacy protection offered to end-users.
Overall, our findings are as follows:

\begin{enumerate}[label=(\roman*),itemsep=0pt] 
\item We show that \toolname achieves an efficient trade-off between user anonymity and estimation precision for encrypted RTB prices.
\item We show that desktop RTB prices have grown $4.6\times$ over desktop RTB prices measured in 2013.
\item Compared to mobile RTB prices measured in 2015, desktop RTB prices have grown by a factor of $3.8\times$. 
\item The advertising value of women appears to be $2.4\times$ higher than that of men.
\item In terms of age, 25-34 year old users appear to be $2.5\times$ more valuable to advertisers than 35-44 year users.
\item Different age and gender groups appear to receive varying attention (and targeting) by different DSPs.
\item Turning to time-of-day and day-of-week correlations with RTB prices, we see that early mornings and weekends, is when the most expensive ads get rendered. 
\item Finally, we see that if prior to the rendering of an ad, exchange of information has taken place between advertisers and/or data brokers via \emph{cookie-syncing}, ads end-up having a 19\% higher median value than when there's no cookie-sync taking place before.    
\end{enumerate}

\section{Background on RTB}
\label{sec:background}
A Real-Time Bidding (RTB) auction is a programmatic, instantaneous type of auction, where a publisher's advertising inventory is bought and sold on a per ad slot basis.
During such  an auction, advertisers place bids for an ad-slot in a publisher's website and the one with the highest bid gets to render its ad on the user's display.
RTB is more efficient in terms of revenue than the traditional static ad-buying for both the advertisers and the publishers.

\noindent{\bf A typical RTB auction:}
A typical transaction for an ad-slot begins with a user visiting a website.
This triggers a bid request from the publisher (or Supply Side Platform (SSP)) to an Ad-Exchange (ADX), usually including various pieces of user's data (\eg interests, demographics, location, cookie-related info, minimum acceptable price, \etc).
Then, multiple Demand-Side Platforms (DSPs) programmatically submit their impressions and their bids in CPM (\ie cost per thousand impressions~\cite{cpm}) to the ADX.
All bids are sealed so every participant places only one bid for a particular ad-slot; this allows the RTB auction to finish within milliseconds (the entire RTB protocol usually runs in around 100 ms).
The ad slot goes to the highest bidder and its impression is served in the user's display.
Typically, the charge price for the ad slot is the second higher price following the Vickrey auctions~\cite{vickrey1961counterspeculation}.

\noindent{\bf Charge price notification:}
When an ADX selects the winning bid of an auction, the corresponding bidder must be notified about its win and the price to be paid to the ADX.
This happens with a notification message (or nURL) conjoined with the price, piggybacked in the ad-response.
This nURL passes through the user's browser and acts as a call-back to the DSP.
This ensures the DSP that the winning impression was indeed delivered (the callback is fired soon after the impression is rendered on the user's device), and also gives  the opportunity to drop a cookie on the user's device.
The nURL includes the winning DSP's domain, the charge price, the impression ID, the auction ID and other relevant logistics (see example in Table~\ref{tbl:nurls}).
In this work, we monitor such nURLs and we study the prices embedded in them, as well as how they associate with the users' browsing behavior and other personal information.

\begin{table}[t]
	    \caption{Real world example of a RTB price notification URL with the ADX of MoPub and the winning DSP of PocketMath. True IDs are omitted for space efficiency.}		\vspace{-0.2cm}
	{\footnotesize
		\begin{tabular}{l}
			{\bf Example of Winning Price Notification URL} \\
			\toprule
				{\bf cpp.imp.mpx.mopub.com/imp}?ad\_domain=mobileacademy.com\& \\
				auction\_time=1420072833\&\textbf{\textcolor{heraldBlue}{bid\_price=1.17}}\&bidder\_id={...}\& \\
				bidder\_name=PocketMath\&campaign\_id={...}\&\textbf{\textcolor{heraldBlue}{charge\_price=0.95}}\& \\
				request\_id={...}\&response\_id=1420072832890\&units=0\&adgroup\_type= \\ 
				marketplace\&adgroup\_priority=9 \\
				\bottomrule
		\end{tabular}
	}
	\label{tbl:nurls}
\end{table}

\noindent{\bf Estimation of encrypted charge prices: }
Major ad companies like DoubleClick, OpenX, RubiconProject and PulsePoint tend to encrypt important information in the nURL (\eg the charge price) to ensure the integrity of the reported values and avoid the prying eyes of competitors.
Olejnik et al. in~\cite{lukasz2014selling-privacy-auction} assumed that the encrypted prices are following the same distribution with the cleartext ones.
After observing an increase on the number of ad-companies that were using encryption, Papadopoulos et al. in~\cite{imcRTB2017} designed a methodology to estimate the value of encrypted prices showing that encrypted ad prices are (1.7$\times$) more expensive than cleartext.

\noindent{\bf Header Bidding (HB):}
In the last few years, a new advertising standard has been proposed, called Header Bidding~\cite{header_bidding}.
In this protocol, the ad-auction takes place in the user's browser with the help of a javascript library), instead of a remote ADX as in RTB.
Advertising partners receive bid requests from the browser library for an ad slot, and respond directly with their bid.
The publisher (via the library in each user) is therefore informed of all bids and can choose the highest bid to sell the impression.
In contrast to RTB, in HB the publisher has full control of the auction, and knowledge of the advertising partners and bids they place.

\section{System Design}\label{sec:architecture}

In this section, we elaborate on the guiding principles behind the design of the system, the various components it needs and the functionalities they need to execute to compute correctly this cost, while delivering a proper end-user experience.

\subsection{Guiding Design Principles}

We group the design principles into three classes: user-related, system-related and ad-related.
The \textit{user-related} principles are summarized as follows:

\begin{enumerate}[label=(\roman*)] 
	\item \textbf{Simple to understand.} The user interface (UI) of the tool should display metrics and visuals that are simple to understand by users who are neither tech- nor privacy-savvy.
	\item \textbf{Unhampered user experience.} The tool should be easy to install and not degrade the user's existing Web experience.
	It should not change a webpage's appearance, it should not slow down the loading of a webpage, should not affect the network traffic induced by a webpage, and should use a minimum of computing resources on the user's device.
	\item \textbf{Preserve user anonymity.} The tool should not, in any case, leak sensitive information that could help an entity to re-identify the end-user on the Web.
\end{enumerate}

\noindent The \textit{system-related} principles are summarized as follows:

\begin{enumerate}[label=(\roman*)] 
	\item \textbf{Transparency.} The tool's operations and functionality should be transparent, and an end-user or engineer should be able to audit the tool and its functionalities.	
	\item \textbf{Scalability.} The tool's architecture should be able to scale to accommodate activity delivered by thousands of users.
	\item \textbf{Fault tolerance.} The tool should continue to function even after encountering unexpected behaviour, without the need of user interaction, or causing issues to the user's browser.
	\item \textbf{Privacy-by-design.} The tool should allow users to opt-in to reporting metadata from ads detected, but in a privacy-preserving fashion.
	That is, the data transmitted from the tool should not expose the user's identity.
\end{enumerate}

\begin{figure}[t]
	\centering
	\includegraphics[width=.5\textwidth]{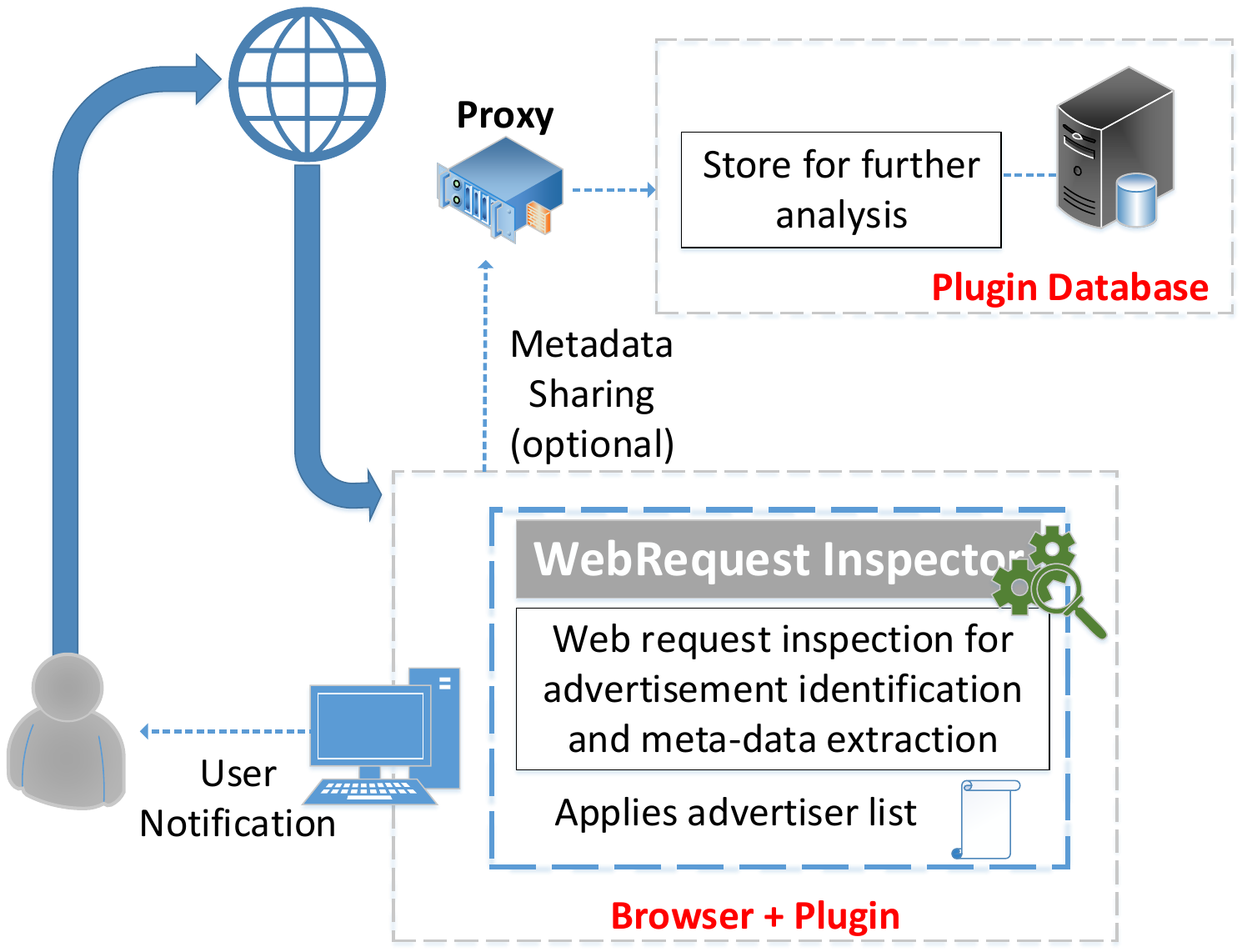}	
	\caption{Overview of the \toolname architecture. 
		The user visits the publisher's website and the extension inspects all the webRequests to find any possible ads.
		When an ad is detected, the extension extracts metadata related to the ad, anonymizes them and, if the user chooses to do so, sends them to a proxy which forwards them to the server, for storage and historical analysis.
		At the same time, the UI incorporates the new price extracted from the ad, and displays an updated total value of the user to advertisers.}
	\label{fig:overview-architecture}
\end{figure}

\noindent The \textit{ad-related} principles are summarized as follows:
\begin{enumerate}[label=(\roman*)] 
	\item \textbf{Real-time operation.} The tool should operate in real time and be able to inform the user of updated measures as soon as new RTB ads are detected.
	\item \textbf{Generality of RTB detection.} The tool should detect RTB ads regardless if delivered in regular webpages (e.g., news) or closed-up publishers (e.g., social media).
	\item \textbf{Untampered RTB ad-protocol.} The detection mechanism should be simple and not tamper with the RTB protocol responsible for the impressions logistics, nor add on its expected latency to operate correctly.
	\item \textbf{RTB price detection.} The detection mechanism should detect and consider the value for both encrypted and cleartext prices from RTB nURLs.
\end{enumerate}

\toolname satisfies all aforementioned principles.
It provides a simple UI to its end-users that reports in real time RTB information without altering the user experience or the RTB protocol, and these two aspects were confirmed after extensive testing.
At the same time, its development process was guided by security and privacy-by-design principles to guarantee security and anonymity of online users.
\toolname is an open source plugin and its code is available in our Github repository.
More information can be found in the plugin's webpage hosted at https://youradvalue.tid.es:2222.

\subsection{System Components \& Functionalities}
Following these guidelines, we design the system and its various components.
In particular, the system has two main components:
1) the end-user client (web browser extension),
2) the web server (running behind a web proxy).
Figure~\ref{fig:overview-architecture} illustrates an overview of the system's architecture.
In the next paragraphs, we explain each of the components and how they fulfill the listed properties.
The extension is written in a few thousand lines of JavaScript and is available for all major browsers.
Additionally, the extension can be inspected at the user side while running on the browser.

\subsection{User Interface \& Input}
\begin{figure}[t]
	\centering
	\includegraphics[width=.2\textwidth]{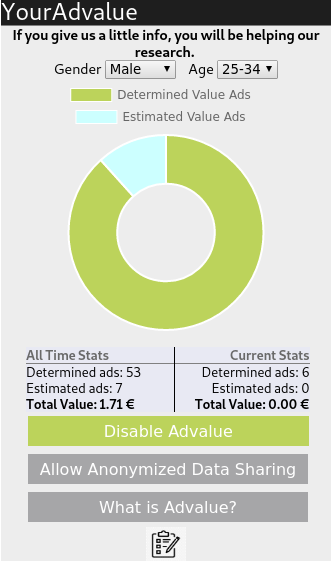}
	\caption{The user interface in the popup that the user gets while using \toolname browser extension. It includes:
		(i) gender and age fields which are user-specified and optional,
		(ii) pie chart which displays the type of ads a user encounters in a nice format,
		(iii) statistics field, where all time and session information about the ads are displayed, and
		(iv) menu, where the user can tune the extension.}
	\label{fig:ui-example}
\end{figure}

For the end-user client to adhere with the user-related principles, we opt for building a browser extension which allows easy access to the user's browsing sessions, can provide an easy to use UI, and can perform network traffic inspection without interfering with the normal operation of the website under visit.
The extension executes a web request inspector, which analyses the incoming and outgoing network traffic of a website and detects RTB-related requests (nURLs).
An example of this user interface (UI) can be seen in Figure~\ref{fig:ui-example}.

This UI is simple and intuitive and was designed with the help of UX experts.
In this UI, the user is prompted to indicate (if they want) their gender and age range.
Furthermore, it displays to the end-user various simple metrics such as the total cost computed for the said user since the tool's installation, as well as the cost of the user for the current session.
The RTB ad detection and ad-related metadata extraction methods are described next.
The design of the extension is light and its operation was tested with hundreds of different websites to confirm it does not hinder the normal loading or operation of each site.
Finally, buttons allow the user to switch it on/off, and opt-in/out at any time from anonymous data reporting.

\subsection{Web Server \& Database}
At the backend, the server receives the anonymous metadata reported for further analysis.
Upon arrival, the data are shuffled  (reordering the table to remove contiguous rows from the same user) with the already collected data to break any relationship of reported data with their reporting users.
The collected data are subsequently cleaned and analyzed for retraining and updating the decision tree (DT) model for the encrypted prices (as described in~\cite{imcRTB2017}).
The updated DT is sent at frequent intervals to the user clients.
For this transmission, the tree is serialized using a proper and agreed structure (\eg XML format) and deserialized at the browser extension.

\subsection{RTB ad detection \& price extraction}

\noindent\textbf{nURL detection.}
The browser extension analyses all network traffic outgoing from the browser to detect all RTB-related requests.
For each request leaving the browser, the extension collects metadata to check if it is a nURL about an ad.
To detect such requests, we employ the \textit{webRequest} API, which is common and available in all browsers we support.
In particular, we create a listener using the event \textit{onBeforeRequest}, an event which is triggered when a request is about to be made.
The goal is to detect RTB nURLs; we don't want to block or redirect the nURLs and we don't want to tamper with the RTB protocol or affect the user experience.

We note that in RTB, the auction information is opaque to the client and the only information that can be inferred is through the parameters of the nURL.
In contrast, in the HB protocol, browser DOM events are triggered that contain metadata directly available at the user browser.
We use such events to clearly distinguish between RTB and HB-related prices, thus, focusing on the former.

\begin{figure}[t]
	\centering
	\includegraphics[width=.5\textwidth]{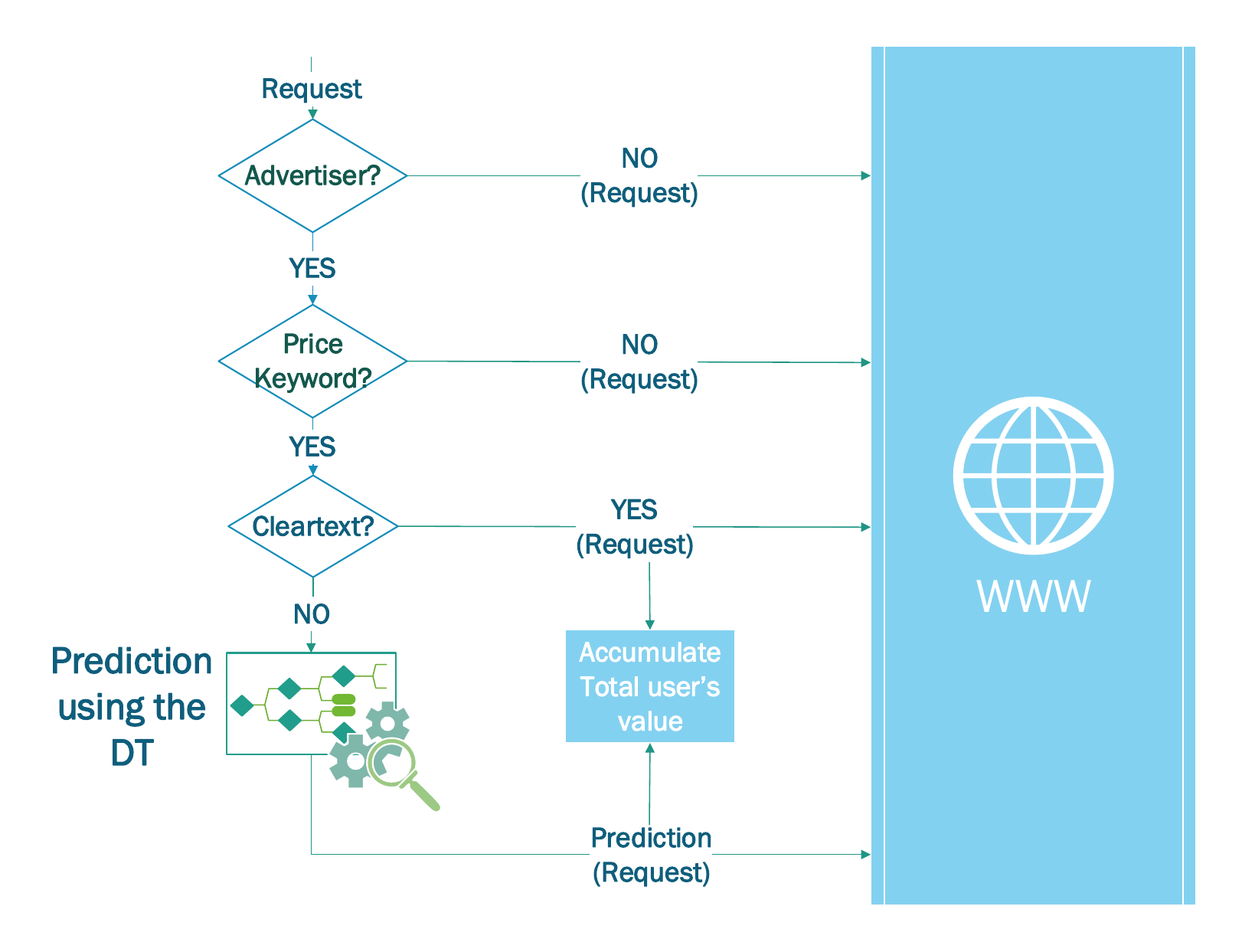}
	\caption{The extension checks if the request corresponds to a known advertiser and if the price is encrypted.
		If it is not a known advertiser, it is let through.
		If it is cleartext, the tool extracts the value of the ad and adds it to the user's total ad-cost.
		If it is encrypted, the tool applies a decision tree model to infer its value before adding it to the user's total ad-cost.}
	\label{fig:request_flow}
\end{figure}

\noindent\textbf{DSP matching.}
After the tool catches an outgoing request, it checks if it is a nURL or not, by comparing the destination to a set of known DSPs (Figure~\ref{fig:request_flow}).
If it does not match with any of the DSPs, it is allowed to pass.
If it is, we consider it a potential notification URL and extract the metadata related to it (i.e., all http variables and values accompanying them).

\noindent\textbf{Price extraction.}
We then check if there is a possible price keyword in the nURL's metadata, and if this keyword is correlated to the nURL's advertiser (we utilize lists of keywords provided by~\cite{lukasz2014selling-privacy-auction, imcRTB2017}).
The argument here is that each advertiser or DSP is associated with a specific keyword used in the RTB.
Therefore, if the price keyword does not match the typical keyword used by the detected DSP, it is considered a false positive and let through.
However, if the keyword matches what the DSP uses, the tool extracts the associated price value.

\noindent\textbf{Encrypted vs. cleartext prices.}
By leveraging the methodology of~\cite{imcRTB2017}, the extension is able to detect both cleartext and encrypted RTB prices (\ie if the price is numeric or alphanumeric).
If the value is numeric (cleartext), it is normalized to CPM in US dollars and added to the user's total ad-cost.
Otherwise, the tool applies the provided Decision Tree (DT) to estimate the value, and then add it to the total ad-cost.

\noindent\textbf{Encrypted price inference with DT model.}
Classifying an encrypted price from a nURL is not a trivial task.
First, before the tool even detects a nURL, it needs to load the DT model provided by the web server.
The model is constructed at the server and exported as an XML file which is rolled out with the extension.
The extension parses this XML file and creates an internal representation of the model that the server has created.
We choose to embed (or preload) the DT model in an XML representation inside the extension for two reasons:
(i) the DT does not change frequently and an existing version can be preloaded, and
(ii) to reduce the number of requests from the extension to the server to minimum needed.
Whenever the DT is updated from the server (normally every few months, when the server has enough new data to retrain the DT), the extension receives it in the form of an update of about 60-70KB.
For the actual prediction of the encrypted price, we implemented a lightweight and efficient function that parses the DT and outputs the price prediction, so it does not interfere with the user experience, nor slow down the user's browsing (each price prediction takes less than 1ms).
If the extracted feature values cannot be matched to the DT model, the extension uses a rolling time average of the cleartext prices as an estimation of the detected encrypted price that could not be inferred.

\noindent\textbf{Feature extraction.}
When an encrypted price is detected, the tool extracts the features that the DT requires to make the price inference.
For the feature extraction process (explained next), we collect metadata about the publisher and the detected ad.
These features are provided to the prediction function to perform the inference.

\noindent\textbf{Extension overhead.}
We tested our extension and did not notice any impact in the user experience.
The analysis of a nURL takes less than 1ms, even when the price is encrypted.
Also, when the user allows it, data sharing happens in an asynchronous manner which does not affect the page load time, or the total available bandwidth.

\subsection{Ad-related feature extraction}
\label{sec:features}

\noindent\textbf{Location extraction.}
For the location extraction, we use an online API.
By making a GET request over a secure channel (HTTPS), the extension obtains and stores the user location on a country level.
The location is stored in the client for the whole session (as long as the user keeps the browser open, does not restart the extension, \etc), reducing requests and other network resource needs.

\begin{table}[t]
	\centering
		\caption{Metadata that the data-contributing users send to the server for further analysis and model updates.}\vspace{-0.2cm}
	{\scriptsize
	\begin{tabular}{cll}
		\toprule
		& \textbf{Feature} & \textbf{Feature Description} \\
		\midrule
			& Gender & User's chosen gender \\
		
			& Age & User's chosen age \\
		\multirow{-3}{*}{\textbf{\begin{tabular}[c]{@{}c@{}}User \\ Information\end{tabular}}}
								& Location & User's current location \\  \hline
			& Time Of Day & Time of the day an ad was detected \\
			& Day Of Week & Day of the week an ad was detected \\
			& Category & Topic category of 1st party domain \\
		\multirow{-4}{*}{\textbf{\begin{tabular}[c]{@{}c@{}}Browsing \\ Information\end{tabular}}}
								& DoNotTrack & DoNotTrack flag enabled/disabled \\  \hline
			& Ad Format & Impression's size in display \\
			& Winner DSP & The DSP that won the auction \\
			& Price Keyword & Keyword with the RTB ad-price \\
			& Price Value & RTB price detected in the nURL \\
		\multirow{-5}{*}{\textbf{\begin{tabular}[c]{@{}c@{}}Advertisement\\ Information\end{tabular}}}
		& Cookie Syncing & Cookie Synchronization detected \\
		\bottomrule
	\end{tabular}
	}
	\label{table:features}
\end{table}

\noindent\textbf{Ad-format extraction.}
After analysing several nURLs, we created a list with keywords stating width and height of ads.
Based on that list, we examine the nURLs' parameters for possible matches that are then stored in the price features.

\noindent\textbf{Cookie Synchronization extraction.}
Except from detecting RTB, we are interested in detecting if a user is being tracked by a 3rd-party entity.
To achieve that, we again employ the \textit{webRequest} API and in particular, the \textit{onBeforeRequest} event.
We attempt to detect if user identifiers are sent from the publisher's webpage to other hostnames.
We first load all the cookie identifiers stored in the user's browser, discarding session cookies and those with values less than 10 characters because most of times they have values unrelated to user identifiers.
For each request from a tab, we check if a user identifier is present in the URL, and if that URL is sent to a tracker.
Also, we make sure that the URL detected is not from the same domain that the user is currently visiting in that tab.
Our method is able to detect a user identifier if it is included in the URL's parameter values, or as part of the URL path.
When we detect such tracking from a 3rd-party, we store the information in a binary flag, indicating that user tracking with cookie synchronization took place in a specific tab to a specific domain.

\noindent\textbf{IAB category extraction.}
Websites can be categorized based on taxonomies such as IAB's~\cite{iabCats}, which are well adopted by the ad-ecosystem.
In order to extract the IAB category for each 1st-party, we created a list with a mapping between websites and their IABs.
We choose to include only very popular websites to keep the list small and the extension lightweight.
This list contains the top 500 Alexa sites, and can be updated from the server as frequently as the DT.
When the user visits a website, we extract the 1st-party domain and check the list.
In case we have a match, we report the IAB found, otherwise we report ``undefined''.

\subsection{User- \& ad-related metadata reporting}
\label{sec:metadata}
As mentioned earlier, \toolname allows users to contribute with their data like ad-prices and auction metadata.
In Table~\ref{table:features}, we list the metadata reported to the server.
The list of features is carefully selected to reflect the needed features for the DT modeling to happen at the server.
It also enables further analysis of RTB prices, how they evolve through time and if the advertisers target particular user categories based on their demographics (\eg gender, age, \etc).

Being very cautious to preserve the user anonymity and privacy, the data reported do not include any user identifiers or other Personally Identifiable Information (PII).
Also, they are anonymized at the client side before being sent to the server.
Each feature can be reported at different granularities.
However, one crucial question in this work is what is a sufficient granularity for a given feature to allow effective price modeling, but not compromise user anonymity?
In Section~\ref{sec:privacy-evaluation}, we elaborate on the aggregation or other obfuscation methods via noise addition that are possible for the given features, and the guarantees or limits on anonymity they can offer to the reporting users.

Also, in order to overcome the problem of an honest but curious server trying to re-identify the sender, each extension reports its data at random times throughout the day and week.
Because of this induced randomness, reports may delay from several seconds to days after they are created, before leaving the user's browser.
Therefore, the server cannot identify a user based on the rate of its reports, since reporting from different users gets mixed-up while being received at the server side.
In Section~\ref{sec:privacy-evaluation}, we also investigate possible threats on anonymity of users reporting such data, and how the system's design can protect them from such de-anonymization attacks.

\section{Dataset Collection \& Limitations}
\label{sec:dataset}

\toolname is designed with user anonymity and privacy-by-design in mind.
Thus, we need to ensure that all contributing users cannot be (re)identified or correlated with their reported data, and that user profiles cannot be constructed.
To keep that promise, we do not store user PII, and do not make attempts to correlate user data shared with the server.
This comes with the limitation that we cannot know exactly how many users contribute data, and with what rate, \ie when and what is reported, and by whom.

To attract users for our study, we advertised the \toolname plugin across different social media networks and groups, and received interest from users across different demographics and communities.
Since we made our plugin available for download from the Chrome store \footnote{https://chrome.google.com/webstore/detail/youradvalue/apipekdgpjmbooidaohhecfpaecahaap}, we were able to get an intuition of the active users from the information page that Google provides to the developers, which indicates the active users at each time.
We observed that the rate of data sharing followed a similar trend to the active users of our plugin.
That is, while the number of active users (\ie that installed the plugin) increased, so did the rate of data sharing.
Similarly, when users removed our plugin or deactivated it, the reporting rate dropped.
Overall, a total of \rusers users agreed to download and install our extension in their primary browser, providing us with real user data, but also imposing some limitations as well.

First, the users were free to use ad-blocking software.
Disabling the ad-blocker is not obligatory for the tool to work, but we advised the users not to use ad-blocking, so that they see the tool in action and receive a better estimate of their ad-cost.
This is because ad-blocking reduces the number of ads delivered successfully to the user browser. 
Therefore, RTB ad detection on the browser, as well as data reporting to the server, can become sparse or even non-existent, depending on the effectiveness of the ad-blocker in place.

Second, the number of users was not stable, since they were allowed to install the plugin, and then enable it and disable it at will, or even uninstall it completely.
Therefore, the number of \toolname users we report (\rusers) is the distinct users that installed the extension at some point during the data collection, and likely contributed with some data.

Third, the reporting users are assumed to be real users, since they need to have a Google account, and proceed to several manual actions such to: search, find and install the specific plugin (or reach it via the tool's website\footnote{https://youradvalue.tid.es:2222}), disable their ad-blocker if they can (for useful data to be collected), and enable the plugin to report anonymized ad-related data to our server, and optionally, declare their specific demographics (as explained earlier).
We do not have control over any of the above actions the users choose to perform (or not), and therefore, cannot know how many of them are truthful in their data reporting.
However, since \toolname is a plugin related to a specific interest, and was advertised in different privacy-aware communities via social media, we believe our audience has some knowledge of the topic and no intention to be malicious.

Fourth, users of specific demographics may have preference to particular (categories of) websites.
Indeed, we have no control over the websites visited by users, which introduces a limitation on comparing users of opposite-sided demographics (e.g., younger vs. older, or male vs. female users).
In our analysis in the next sections, we assume that the data collected per demographic represent well each user category, and that we can cautiously generalize our findings as such.

\begin{table}[t]
	\centering
	\caption{Summary of the datasets or values available from our work and past studies.}\vspace{-0.2cm}
	{\scriptsize
		\begin{tabular}{lrrr}
			\toprule
			\textbf{Property}	&	\textbf{Desktop ($U$)}&	\textbf{Mobile~\cite{imcRTB2017} ($M$)}&	\textbf{Desktop~\cite{lukasz2014selling-privacy-auction} ($D$)}	\\ 
			& {\bf \toolname} & \emph{\bf Papadopoulos et al.} & \emph{\bf Olejnik et al.} \\
			\midrule
			Time period		&	2019 (11 mon.)		&	2015 (12 mon.)				&	2013 (1 mon.)	\\
			Real users		&	$\sim$\rusers		&	$\sim$810					&	$\sim$100		\\
			RTB Ads detected	&	11946			&	78561					&	$\sim$400		\\
			\rowcolor[HTML]{EFEFEF}
			Gender			&	Yes				&	No						&	No			\\
			\rowcolor[HTML]{EFEFEF}
			Age				&	Yes (5 bins)		&	No						&	No			\\
			\rowcolor[HTML]{EFEFEF}
			Location			&	10 countries		&	Spanish cities				&	3 countries	\\
			Time Of Day		&	Yes (8 bins)		&	Yes (8 bins)				&	Yes (3 bins)	\\
			Day Of Week		&	Yes (7 days)		&	Yes (7 days)				&	No			\\
			\rowcolor[HTML]{EFEFEF}
			Cookie Synching	&	Yes				&	No						&	No			\\
			\rowcolor[HTML]{EFEFEF} 
			DoNotTrack		&	Yes				&	No						&	No			\\
			\rowcolor[HTML]{EFEFEF} 
			adFormat			&	Yes				&	Yes						&	No			\\
			Winner DSP		&	Top 20 entities		&	Top 15 entities				&	Top 25 entities	\\
			IAB category		&	30 IAB categ.		&	30 IAB categ.				&	14 Web categ.\\
			\bottomrule
		\end{tabular}
	}
	\label{table:datasets}
\end{table}

\noindent
\textbf{Dataset breakdown:}
The data collected from said users are summarized in Table~\ref{table:datasets}.
As explained earlier, by design, we are not able to identify which users contribute and with how many impressions.
However, we can provide a breakdown of the demographics vs. impressions reported.
In particular, we received 4024 impressions from male and 6640 from female users.
Also, we received 9826 impressions from the 25-34 age group, 1190 from the 35-44 age group.
Finally, we received the following impressions per country: 6600 from US, 2685 from Spain, 1278 from Switzerland, 146 from Greece, 114 from Austria, 24 from Cyprus, 13 from Italy, 6 from Belgium, 4 from Germany and 3 from Israel.
Differences from the totals reported in Table~\ref{table:datasets} are due to users choosing not to report all demographics (\eg they report gender but not age and vice versa), the returned country having the value ``unknown'' because of issues with the location API, \etc

\noindent
\textbf{Comparison with past datasets (Table~\ref{table:datasets}):}
First, we use statistics reported in a past study on RTB in desktop~\cite{lukasz2014selling-privacy-auction} in 2013 (no dataset released) for 100 users and various automated crawls ($D$).
Second, we use a previously collected and shared dataset from~\cite{imcRTB2017} on RTB in mobile, for 2015-2016 ($M$).
This is a larger dataset, containing 1 year-long activity of RTB ad metadata from 2015-16, with about \datasetawazza ads from 810 mobile users, and each entry containing several features.
The last dataset is the aforementioned $U$, with data from real users who installed the \toolname.
This dataset consists of about \datasetyouradvalue ad impressions detected from \rusers users for a period of \toolAvailable.
We note that 64.7\% of the ads collected have cleartext values and 35.3\% encrypted.
We also note that the $M$ dataset has multiple times more ads reported than in the $U$ dataset.
This is because it is constructed from a traffic log of about $4-5\times$ more users, and collected at a Web getaway proxy, instead of using a plugin, thus allowing them to detect all possible RTB-related ads delivered to these users.
However, \toolname collects similar metadata to the ones in the $M$ dataset, so we can directly compare the price distributions for the two time periods, for mobile vs. desktop.
Also, in contrast with previous datasets, we collect anonymized demographics from our users to study their association with RTB prices, something not done before in the past studies~\cite{lukasz2014selling-privacy-auction} and~\cite{imcRTB2017}.

\section{RTB Targeting vs. Demographics} 
\label{sec:rtb-demographics}

\begin{figure*}[t]
	\centering
	\begin{minipage}{0.48\linewidth}
		\centering
		\includegraphics[width=.98\linewidth]{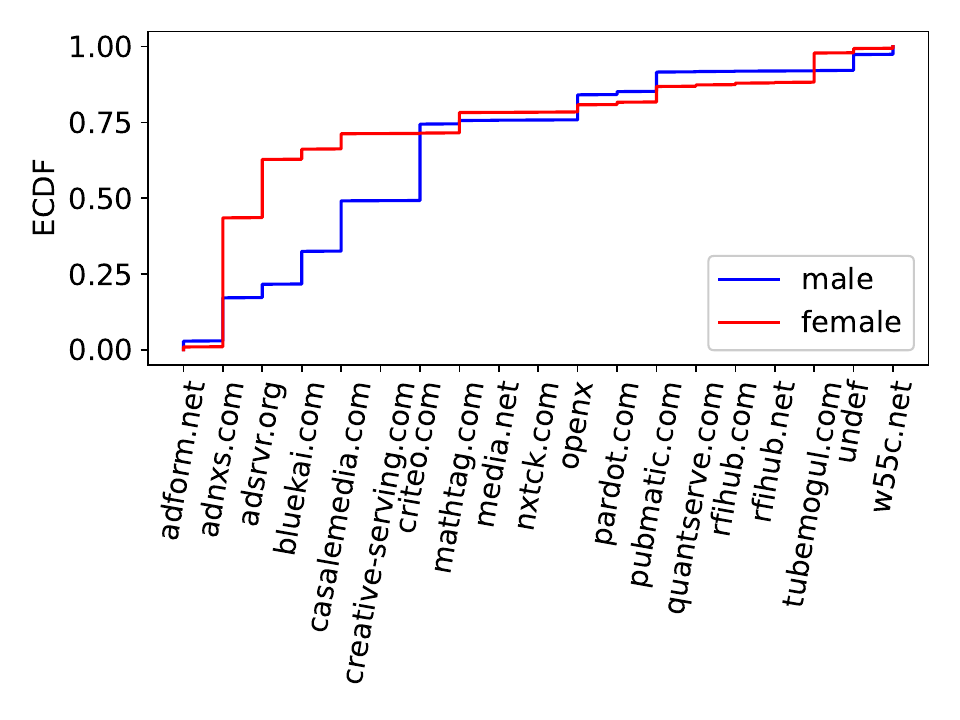}
		\caption{Portion of ads per DSP, per gender.}
		\label{fig:gender_dsp}
    \end{minipage}
	\hfill
	\begin{minipage}{0.48\linewidth}
		\centering
		\includegraphics[width=.98\linewidth]{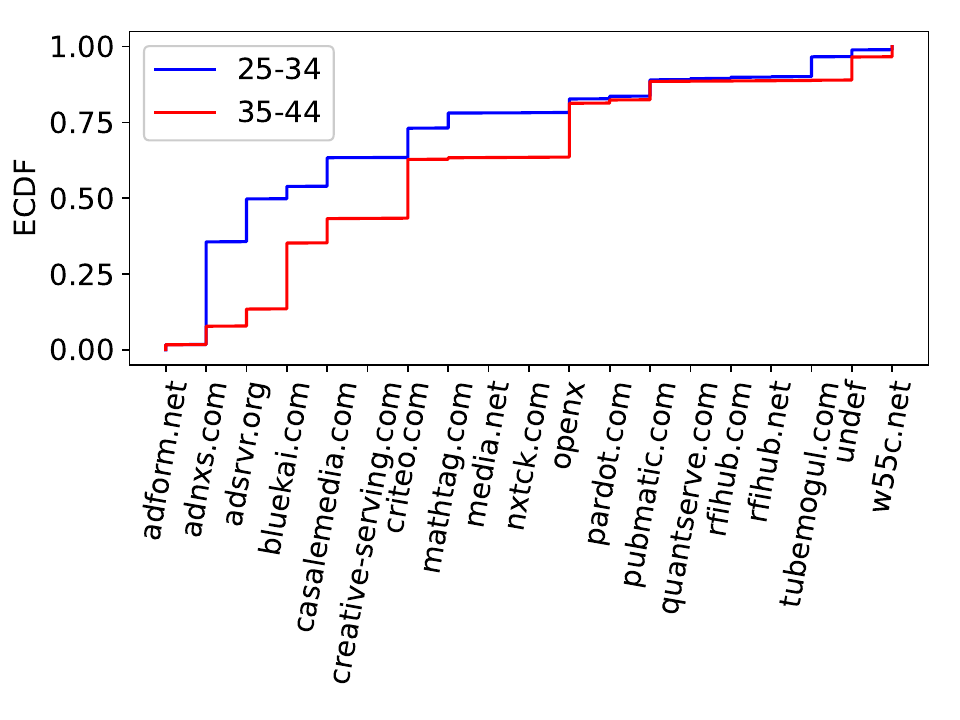}
		\caption{Portion of ads per DSP, per age bin.}
		\label{fig:age_dsp}
    \end{minipage}
\end{figure*}

In this section, we take a first look at data reported by $\sim$\rusers real users who installed the \toolname browser extension, in 2019 ($U$).
Please see Section~\ref{sec:ethics} for details on ethical collection of data.
Table~\ref{table:datasets} summarizes the data and statistics available for this dataset.

In the next paragraphs, we focus on answering the following research question: How do the different demographics (\ie gender, age and location) of online users affect their ad-cost?
In particular, we focus on gender and age, and investigate if a specific age-gender group is being targeted more intensely by the RTB ecosystem (Section~\ref{sec:gender-age-targeting}).
Similarly, we study this question from the point of view of user location, at a country level (Section~\ref{sec:country-targeting}).
Importantly, we proxy the intensity of targeting by studying 2 key elements of RTB: cookie synchronization (CS) during ad delivery, and the corresponding DSP delivering the ad.

\subsection{Gender and Age Targeting}
\label{sec:gender-age-targeting}

First, we assess how the different genders are targeted, by examining the portion of CS detected in RTB ads in conjunction to the gender of the user.
We find that female users received ads with CS in 99.1\% of the time, in comparison to 86.2\% of ads with CS for male users.
Thus, female users are targeted with 12.9\% more CS than males.
Similarly, we assess how different age groups are targeted, by examining the intensity of CS.
We find that users of age 25-34 years received ads with CS in 96.2\% of the time, compared to 79.3\% of ads with CS for the older age group (35-44).
Thus, younger users are targeted with 16.9\% more CS than older users.

Second, we look into genders and the DSPs involved in ads.
Figure~\ref{fig:gender_dsp} shows the CDF of 18 different DSPs involved, broken down by gender.
We find that particular DSPs have preference for specific genders.
For example, \textit{adnxs.com} and \textit{adsrvr.org} delivered many more ads to female than male users.
In contrast, \textit{casalemedia.com} and \textit{criteo.com} had a preference for targeting male users.
Similarly, we study age group and the DSPs involved in ads.
Figure~\ref{fig:age_dsp} shows the CDF of the DSPs involved, broken down by age.
We find that particular DSPs have preference for specific age.
For example, \textit{adnxs.com} and \textit{adsrvr.org} delivered more ads to younger users (25-34).
In contrast, \textit{bluekai.com}, \textit{criteo.com}, and \textit{openx.com} had a preference for targeting older users (35-44).
Interestingly, and as we will see in the next section, this variability or preference in targeting per gender and age, can potentially impact the RTB prices of ads delivered.

\noindent\textbf{Finding:} Female users are targeted more compared to male. Similarly, younger users are targeted more than older users. DSPs may have a preference for specific genders and age groups.
This can typically be the result of the type of campaigns they were running at the time of data collection.

\subsection{Country Targeting}
\label{sec:country-targeting}

\begin{figure*}[t]
	\begin{minipage}{0.48\linewidth}
		\centering
		\includegraphics[width=.9\linewidth]{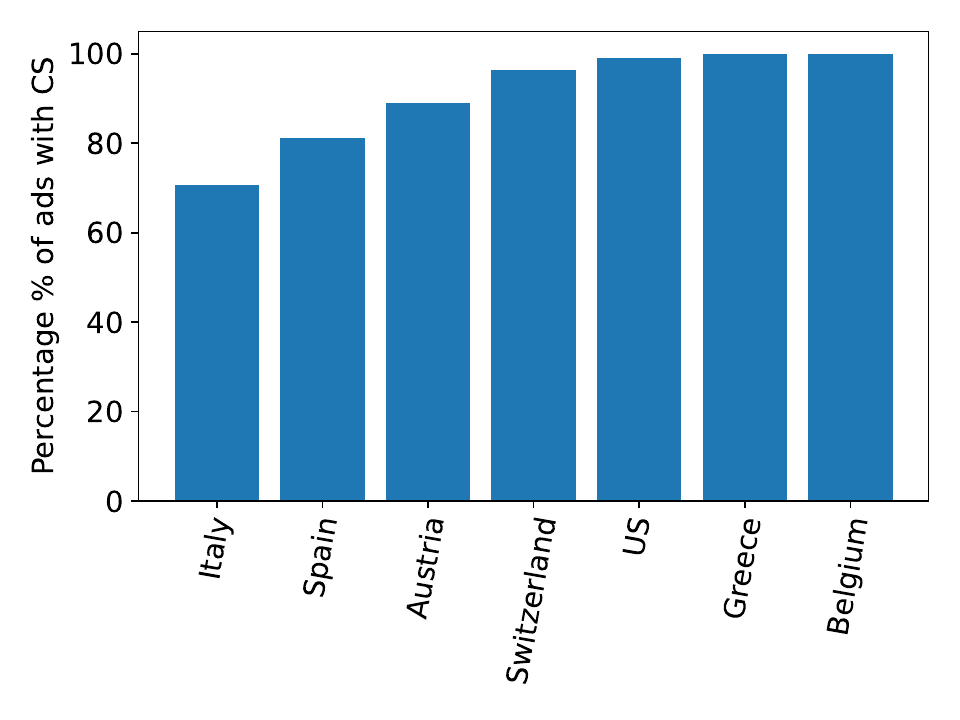}
		\caption{Portion of ads with CS per country.}
		\label{fig:country_cs}
    \end{minipage}
	\hfill
	\begin{minipage}{0.48\linewidth}
		\centering
		\includegraphics[width=.9\linewidth]{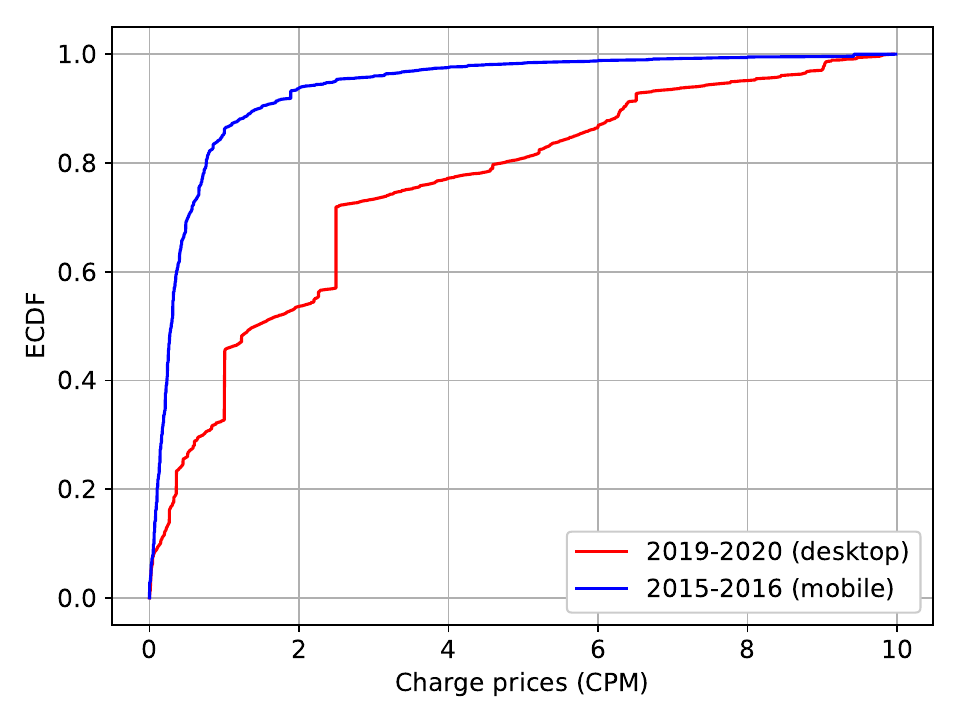}\vspace{-0.2cm}
		\caption{Price distribution for $M$ (2015-2016) and $U$ (2019-2020).}
		\label{fig:price-comparison-old-new}
    \end{minipage}
\end{figure*}

We finish this investigation by comparing the ads delivered to users of different countries, with respect to the CS they received.
Figure~\ref{fig:country_cs} shows for top 7 countries a breakdown of the percentage of ads delivered to users in these countries, that contained CS.
We find that all countries had high portion of ads delivered with CS involved, i.e., 70.6\% or more.
Unsurprisingly, US as the most advanced ad-market, has one of the highest CS portions, along with smaller countries such as  Belgium, Greece and Switzerland.

\noindent\textbf{Finding:} US users are targeted more than EU users. This could be explained, since US has a more mature and aggressive advertising system, and still a more relaxed online privacy regulation framework compared to EU.

\section{RTB Price Evaluation}
\label{sec:price-evaluation}

In this section, following the analysis of demographics, we perform a price analysis and observe how the RTB ad-prices change over time, and how characteristics of users may be correlated to these prices.
To do this analysis, we use all available results reported in literature on RTB prices.
In particular, we aim to answer the following research questions:
\begin{enumerate}
    \item How have RTB ad-prices changed over time? (Sec.~\ref{sec:trends})
    \item Which day of the week, and hour of the day, do ads cost more? (Sec.~\ref{sec:day-hour-cost})
    \item For which IAB website category do ads cost more? (Sec.~\ref{sec:iab-cost})
    \item For which gender, age, and country groups do ads cost more? (Sec.~\ref{sec:gender-cost}, Sec.~\ref{sec:age-cost}, and Sec.~\ref{sec:country-cost})
    \item Does the use of cookie synchronization lead to higher costing ads? (Sec.~\ref{sec:cs-cost})
    \item Which DSPs pay more to reach users? (Sec.~\ref{sec:dsp-cost})
\end{enumerate}

As a small suggestion, it might be nice to include some summary statistics to confirm these assertions, as well as a pointer to the section(s) that have further details. 

\subsection{General RTB price trends \& changes}
\label{sec:trends}
In Figure~\ref{fig:price-comparison-old-new}, we compare the CDFs of RTB cleartext prices for the two time periods we have detailed data: mobile $M$ data~\cite{imcRTB2017} and desktop $U$ data.
We observe that the median price in $U$ is $1.42$ CPM (cost per mile, in USD), whereas in the older, mobile dataset it was $0.29$ CPM, pointing to a $3.8\times$ 
increase between mobile and desktop RTB prices in a 4-year period.
Also, the top 5\% of RTB prices of the new dataset $U$ are $\sim$$2\times$ larger ($7.85$ vs. $3.7$ CPM) in comparison to $M$ dataset.
Olejnik \etal in $D$~\cite{lukasz2014selling-privacy-auction} report an average price of $\$0.43$ CPM per user, with a min and max price of $\$0.04$ and $\$1.98$ CPM, respectively.
As explained, we do not collect user identifiers, so we can not analyze prices per unique user.
For this reason, we can only compute the total average price of ads collected across all potential users we had in the data collection period.
We find an average price of \$2.44 CPM (st.dev=\$2.48) per user, which is more than $4.6\times$ 
higher than the average RTB price per user, 7 years ago in dataset $D$. Our results are verified by corporate studies as well, reporting a $4.3\times$ increase~\cite{spendingGrowth,spendingGrowth2} 

\noindent\textbf{Finding:} During a period of 7 years, RTB prices have increased by $4.6\times$. The prices between desktop and mobile also increased by $3.6\times$ in the last 4 years alone.

\subsection{Which day \& hour costs more?}
\label{sec:day-hour-cost}
\begin{figure*}
	\centering
	\begin{minipage}{0.32\linewidth}
		\centering
		\includegraphics[width=1.06\linewidth]{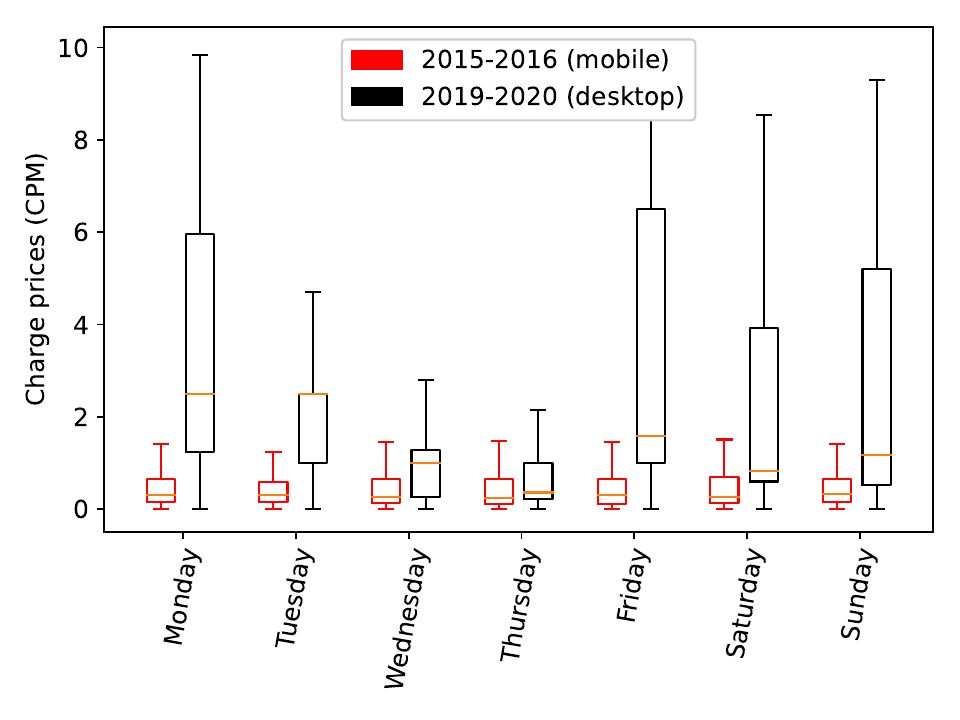}
		\caption{RTB prices per day, for the $M$ (2015-2016) and $U$ (2019-2020) datasets.}
		\label{fig:day_pr}
	\end{minipage}
	\hfill
	\begin{minipage}{0.32\linewidth}
		\centering
		\includegraphics[width=1.06\linewidth]{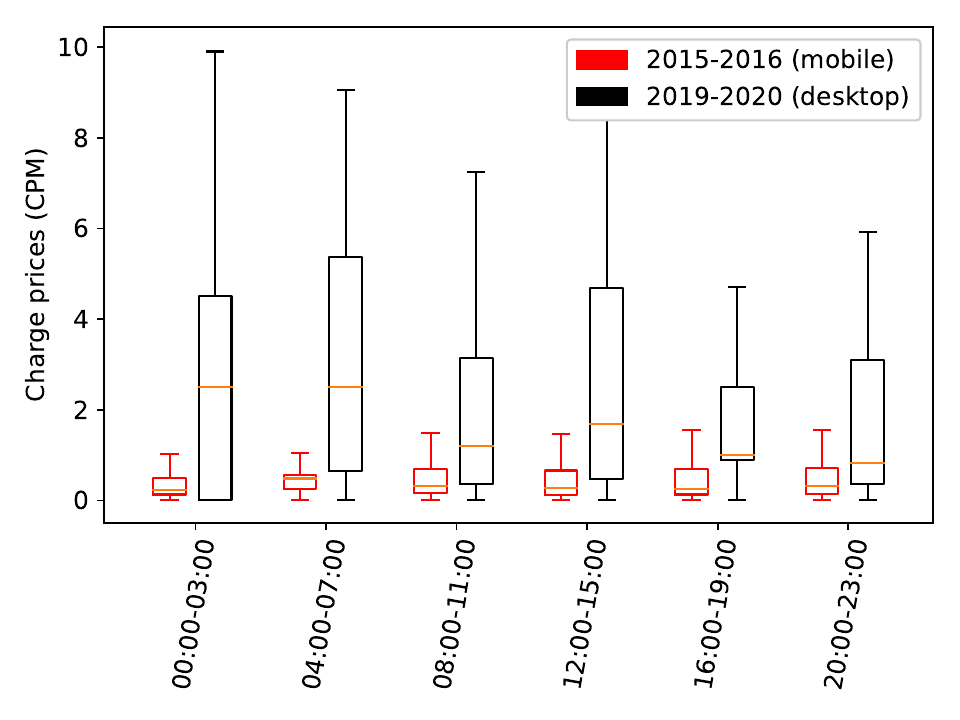}
		\caption{RTB prices per hour slot, for the $M$ (2015-2016) and $U$ (2019-2020) datasets.}
		\label{fig:tod_pr}
	\end{minipage}
	\hfill
    \begin{minipage}{0.32\linewidth}
	\centering
	\includegraphics[width=1.06\linewidth]{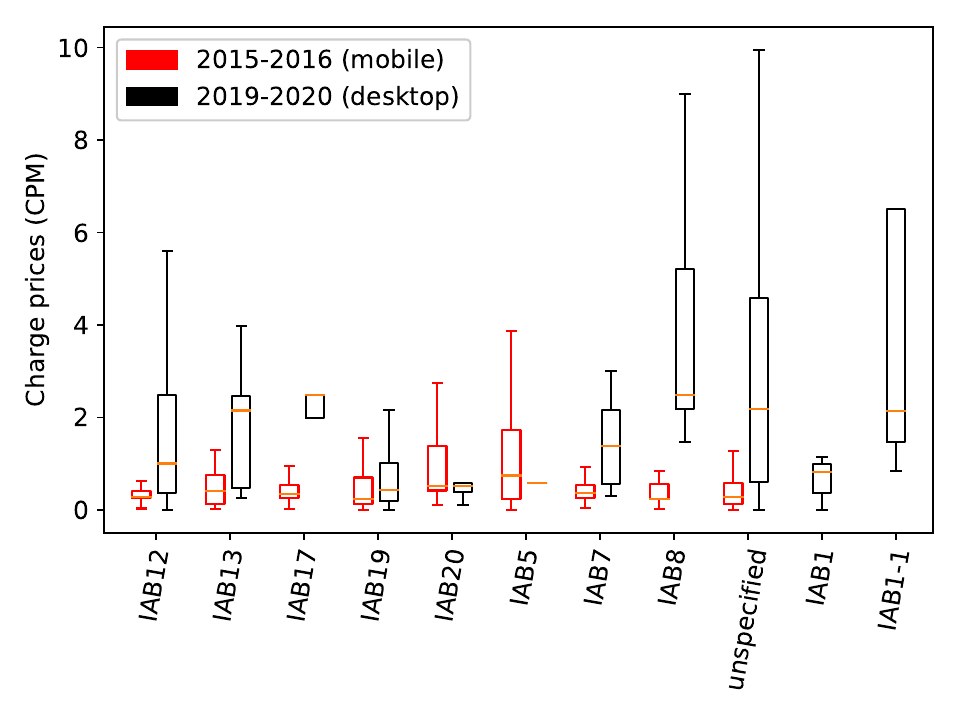}
	\caption{RTB prices per IAB category, for $M$ (2015-2016) and $U$ (2019-2020) datasets.
	}
	\label{fig:iab_pr}
\end{minipage}
\end{figure*}

In Figure~\ref{fig:day_pr}, we compare the prices found for each day of the week.
For $M$, we observe a relatively stable distribution for each day with a median value around 0.3 CPM, and weekdays receiving ad-prices with a higher variability.
In contrast, in the new dataset $U$, we find that during the weekends (Friday-Sunday) and Monday, the ad-prices tend to be higher than weekdays, with medians varying between 0.83 and 2.5 CPM.
Interestingly, on Monday, the median price peaks to 2.5 CPM.
Overall, and acknowledging $M$ is on mobile whereas $U$ is on desktop, we see an increase in prices advertisers pay in 2019 compared to the prices from 4 years ago.

Next, we study ad-prices paid with respect to the time an ad impression was displayed (Figure~\ref{fig:tod_pr}).
To compare our results with the reported from Olejnik \etal~\cite{lukasz2014selling-privacy-auction} 7 years ago, we follow their format and we create 
3 time slots: morning (0-8h), midday (8-16h) and night (16-24h) for the $U$ dataset, and compare with $D$.
As we see, their results confirm our findings: charge prices are higher in the morning slot than afternoon or evening.
However, the prices detected in our dataset ($U$) are in general higher: morning (3.18 CPM vs. 0.57 CPM), midday (2.77 CPM vs. 0.53 CPM) and early night (1.61 CPM vs. 0.47 CPM).

When we compare our findings from $U$ (2019), with the ones from $M$ (2015), we note an increase in 2019 for the morning slot (3.18 CPM vs. 0.65 CPM), in midday (2.77 CPM vs. 0.64 CPM), and night (1.61 CPM vs. 0.67 CPM).
These differences (i.e., when the prices peak with respect to time of day and day of week) can be justified due to different usage of devices: users browse on mobile devices at different hours and days than on desktop~\cite{mobile-and-desktop-usage-1-2017}, and with different intent~\cite{mobile-and-desktop-intent-2013}, thus being delivered ads of different cost~\cite{mobile-and-desktop-ad-cost-2018}.

\noindent\textbf{Finding:} Ad prices are higher on weekends and during the beginning of the week. Also, prices are higher during the beginning of the day. Advertisers are more aggressive during the time users are more relaxed and out of office.

\subsection{Which website category costs more?}
\label{sec:iab-cost}

In Figure~\ref{fig:iab_pr}, we set out to compare the distributions of ad-prices detected in $U$, and how they compare with $M$.
In general, we detected prices across 11 different IAB categories in $U$, compared to 20 in $M$, with 9 overlapping.
We find that the most popular IAB categories in $U$ are News ($IAB12$), Technology and Computing ($IAB19$), and Unspecified.
Similarly, in $M$, the most popular IAB categories are Sports ($IAB17$), Technology and Computing ($IAB19$), or Unspecified.
We also compared the $U$'s IAB categories with the Web categories reported in $D$, if they loosely match, since no IAB taxonomy was used in~\cite{lukasz2014selling-privacy-auction}.
In $D$, the most popular category is News ($IAB12$), followed by Entertainment ($IAB1$), Games ($IAB9$) and Technology and Computing ($IAB19$).

\begin{figure*}[t]
\centering
	\begin{minipage}{0.48\linewidth}
		\centering
		\includegraphics[width=.8\textwidth]{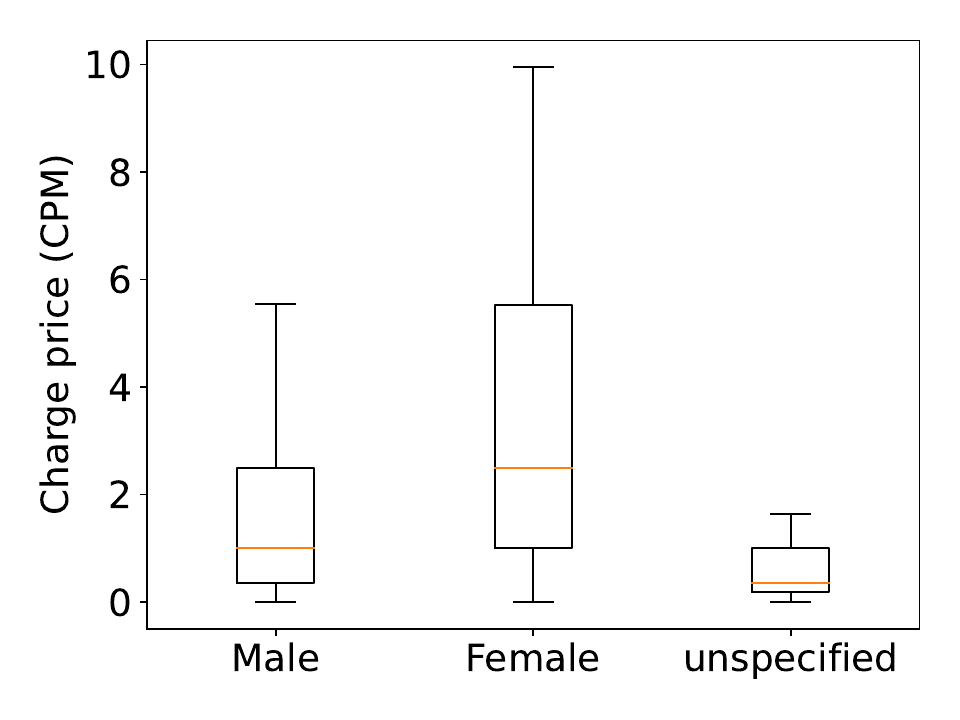}
		\caption{RTB prices per gender, for the $U$ dataset (2019-2020).}
		\label{fig:gender_pr}
	\end{minipage}
	\hfill
	\begin{minipage}{0.48\linewidth}
		\centering
		\includegraphics[width=.8\linewidth]{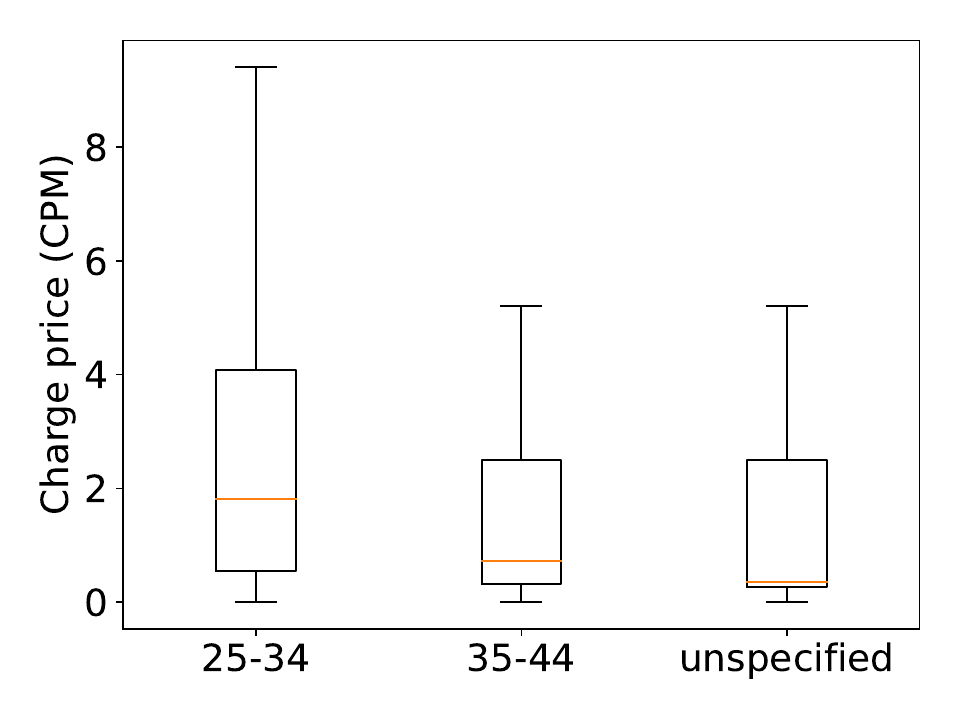}
		\caption{RTB prices by age group, for the $U$ (2019-2020) dataset.}
		\label{fig:age_pr}
	\end{minipage}
\end{figure*}

Using the general statistics provided in~\cite{lukasz2014selling-privacy-auction}, and the distributions of data from~\cite{imcRTB2017}, we compare $U$ with the means of distributions of prices in 2013 ($D$), and 2015 ($M$).
We observe that the average price paid per IAB category of the publisher is not the same across categories.
This has also been reported in~\cite{lukasz2014selling-privacy-auction} and~\cite{imcRTB2017} and it is to be expected, since websites visited by users indicate behavioral preferences and economic status or buying power of the customer. RTB prices in 2019 are higher, on average, for all reported IABs (statistical significance at p-value < 0.001) compared to 2015 and 2013 data.
Comparing $D$ with $U$, the median for $IAB12$ (News) increased from 0.38 to 1.58 CPM, the median for $IAB19$ (Games) increased from 0.38 to 0.82 CPM, and the median for $IAB1$ (Entertainment) increased from 0.33 to 0.88 CPM.
Similarly, when comparing $M$ with $U$, we find the median for $IAB17$ (Sports) to be 0.34 CPM in $M$ vs. 2.5 CPM in $U$, for $IAB19$ (Games) to be 0.24 CPM vs. 0.43, the median for $IAB12$ (News) to be 0.27 CPM vs. 1.0 CPM and in case of Unspecified, 0.28 CPM vs. 2.2 CPM.
Consequently, some browsing interests are more highly targeted by advertisers and drive prices of ads to higher costs.
However, the most popular IAB categories are not the most expensive with respect to ad-cost.
In particular, $IAB17$ (Sports) and $IAB8$ (Food \& Drink) are the most expensive, with median cost 2.5 CPM and 2.49 CPM in 2019, respectively.

\noindent\textbf{Finding:} Widely visited or popular website categories (based on the IAB taxonomy) tend to serve more expensive ads.

\subsection{Which gender costs more?}
\label{sec:gender-cost}

Next, we focus on some demographics of our real user-base from the 2019 $U$ dataset (past studies did not capture these demographics).
In Figure~\ref{fig:gender_pr}, we see that advertisers are willing to pay more to target women than men.
In particular, for female users, the median price was 2.49 CPM, compared to 1.0 CPM for men, and 0.36 CPM for the unspecified.
This can be explained by the earlier analysis on demographics and targeting, which showed that female are more targeted by specific DSPs and with more CS.
This trend confirms past reports (\eg~\cite{menAreCheap}) which claim that women are generally more intensely targeted customers than men, thus, costing more to reach them.

\noindent\textbf{Finding:} Women are $2.5\times$ more expensive to reach with ads compared to men. Advertisers employ more aggressive strategies to target women and are willing to spend more on ads to reach them.

\subsection{Which age group costs more?}
\label{sec:age-cost}
In Figure~\ref{fig:age_pr}, we see the prices paid by advertisers based on the self-defined age of users.
We see that advertisers are willing to spend higher prices for younger users.
In particular, for users in the age group 25-34, the median price was 1.81 CPM, compared to 0.72 CPM for the group 35-44.
Also, for the younger group, advertisers are willing to pay as high as 9.4 CPM, whereas in the older group the maximum reaches 5.21 CPM.
Again, these results for younger users can be explained by higher CS and the tendency of specific DSPs to target this demographic.

\noindent\textbf{Finding:} Advertisers are willing to pay almost double when targeting younger people. Young people could be more receptive to ads, and thus a more profitable audience to target.

\subsection{Which country costs more?}
\label{sec:country-cost}
\begin{figure*}[t]
    \centering
	\begin{minipage}{0.48\linewidth}
		\centering
		\includegraphics[width=.8\textwidth]{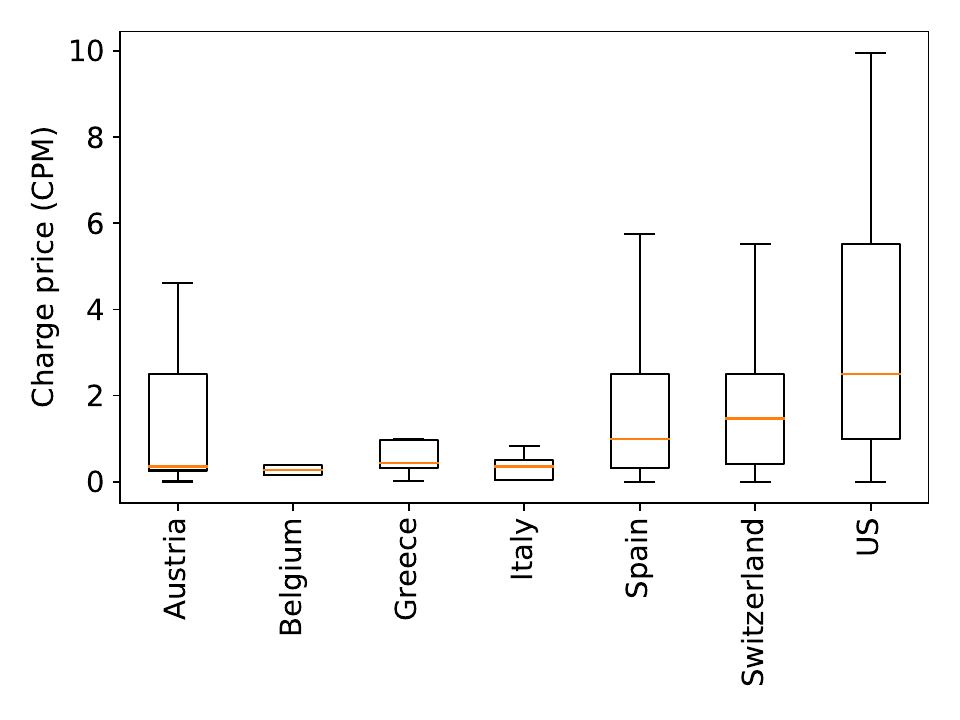}
		\caption{RTB prices per country, for the $U$ (2019-2020) dataset.}
		\label{fig:country_pr}
	\end{minipage}
	\hfill
	\begin{minipage}{0.48\linewidth}
		\centering
		\includegraphics[width=.8\textwidth]{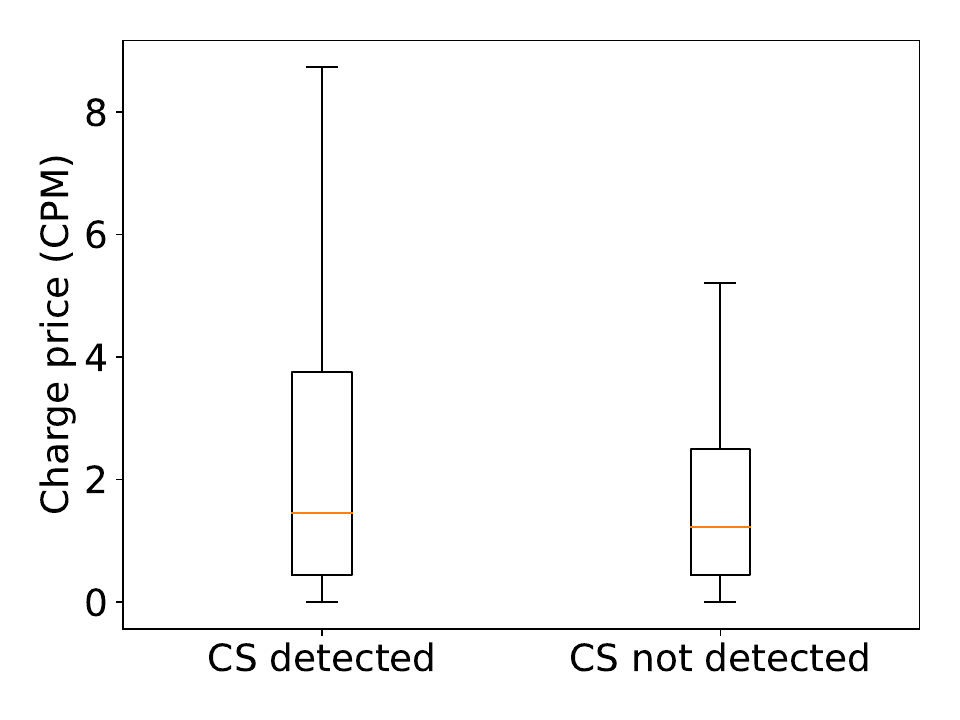}
		\caption{RTB prices vs. CS, for the $U$ (2019-2020) dataset.}
		\label{fig:cs_pr}
	\end{minipage}
\end{figure*}

In general, we detect prices from 11 countries.
In Figure~\ref{fig:country_pr}, we plot the prices for the top 7 by volume.
The top 3 are USA, Switzerland, and Spain, with median ad-prices 2.49 CPM, 1.47 CPM and 1.00 CPM, respectively.
We note that the median price in Spain on the $M$ dataset (mobile) was 0.29 CPM, whereas in $U$ (desktop) is 1.0 CPM.
That indicates an $2.4\times$ increase in CPM in almost 4 years.
Also, in $D$ (2013), the average price in USA was found to be 0.69 CPM, indicating a $3-4\times$ increase with respect to USA ad-prices in $U$ (2019).

\noindent\textbf{Finding:} Users browsing from wealthier countries tend to be more expensive to target. Also, RTB prices increased around $4\times$ in the past 4 years, globally. These results may show a hyperinflation of the specific ecosystem in comparison to the general growth of wealth in countries.

\subsection{Does Cookie Synchronization (CS) indicate more costly ads?}
\label{sec:cs-cost}
CS~\cite{CSync,Acar:2014:WNF:2660267.2660347} is a very important technique for user data sharing, tightly connected with the RTB ad-auctions~\cite{bashirtracing,lukasz2014selling-privacy-auction,papadopoulos2018cost}.
Interestingly, and as shown in Figure~\ref{fig:cs_pr}, the existence of CS is associated with higher RTB charge prices.
In particular, when CS is detected, the median prices are 1.46 CPM, or 19\% higher than when CS is not detected.
In fact, the maximum prices with CS can reach 8.73 CPM, in comparison to 5.21 CPM without CS.
These results support the intuition built by past studies~\cite{bashirtracing}, that when CS is performed before a RTB auction bid, it facilitates user ID syncing between the CS partners, thus enabling user re-identification and retargeting, and a more informed bidding for the DSP.

\noindent\textbf{Finding:} When CS is enabled, the median prices are 19\% higher than when it is not. This verifies the hypothesis that targeted users are more valuable to advertisers compared to new users.

\subsection{Which DSPs pay more to reach users?}
\label{sec:dsp-cost}

\begin{figure*}[t]
    \begin{minipage}{0.45\linewidth}
		\centering
		\includegraphics[width=1.05\linewidth]{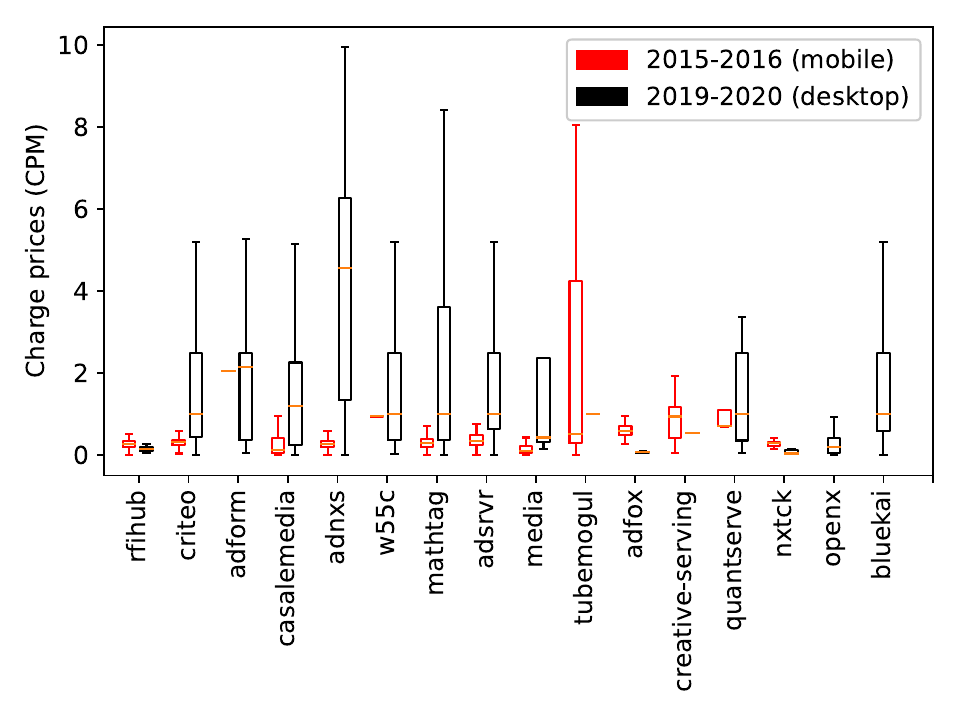}
		\caption{Major DSPs detected, for the $M$ (2015-2016) and $U$ (2019-2020) datasets.}
		\label{fig:dsp_pr}
	\end{minipage}
	\hfill
	\begin{minipage}{0.45\linewidth}
		\centering
		\includegraphics[width=1.05\linewidth]{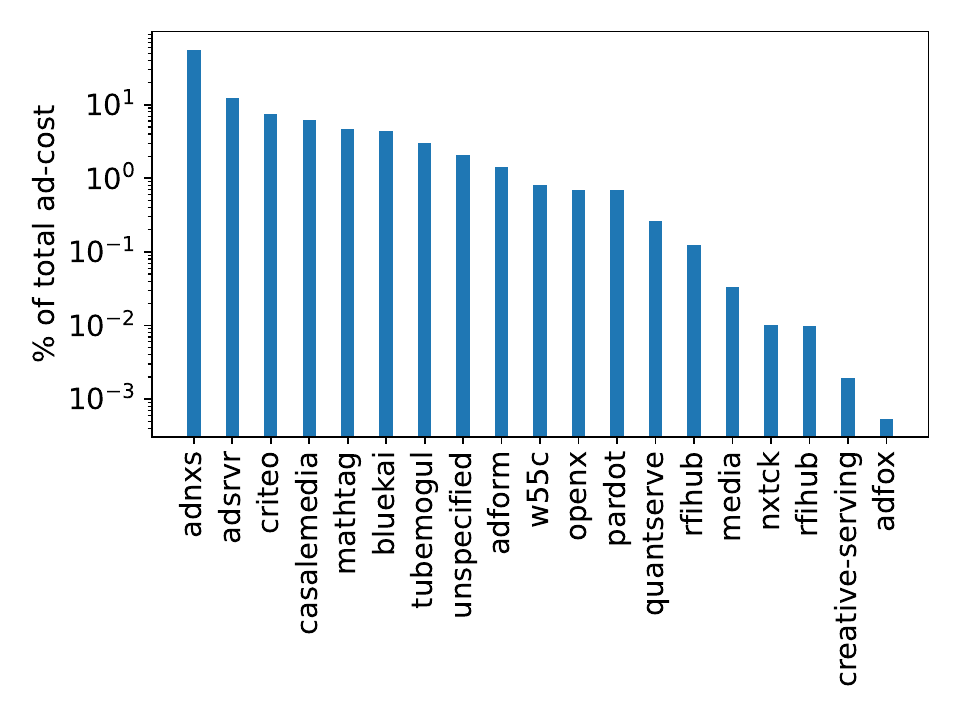}
		\caption{Portion of cumulative spending per DSP, in the $U$ (2019-2020) dataset.}
		\label{fig:dsp_spending}
	\end{minipage}
\end{figure*}

Figure~\ref{fig:dsp_pr} presents a breakdown of the 16 major DSPs involved in RTB for the detected ad-prices.
We find that 6 DSPs are involved with more than 75\% of the ad prices detected in the $U$ dataset.
Interestingly, when compared to the DSPs found in $M$, there are differences which are to be expected, given that $M$ was focused on mobile traffic.
In particular, the top DSPs are $adxns$, $adsrvr$, and $criteo$, with median prices ($U$ vs. $M$): 4.56 vs. 0.27  CPM, 1.0 vs. 0.34  CPM, and 1.0 vs. 0.31  CPM, respectively.
When comparing $U$ to the $D$ dataset, we find that the top DSPs are $invitemedia$, $turn.com$, and $mathtag$.
From those DSPs we were able to detect advertisements only from $mathtag$ with median price ($U$ vs $D$): 1 CPM  vs. 0.09 CPM.
These results demonstrate that the top DSPs have increased the ad-prices they pay to reach desktop audiences by $3-17\times$ through a period of 4-7 years.
Also, in Figure~\ref{fig:dsp_spending}, we show the portion of ad-cost that each DSP spends in RTB ads, in the $U$ dataset.
As expected, we find the top DSPs spend the most in order to reach online users.
In particular, $adnxs$ spends 55.6\%, $criteo$ spends 7.46\%, and $adsrvr$ spends 12.58\% of the total amount spent to reach users.

\noindent\textbf{Finding:} All DSPs increased the prices they are willing to pay for ads in the past 7 years, by a factor of 3-17$\times$.

\section{Privacy Analysis}\label{sec:privacy-evaluation}

In order for the \toolname system to accurately estimate encrypted charge prices, and keep its price modeling engine up-to-date with the current pricing dynamics in the ad-ecosystem, it requires data from the users.
Given the system's principle for privacy-by-design, \toolname needs to protect users' anonymity and privacy against re-identification attacks after a possible data-breach or a malicious server controller.
In this Section, we first elaborate on possible threats that can expose a user's identity, allowing an attacker to link a user to his reported data, and discuss how we solve these threats, thus maintaining user anonymity (Section~\ref{sec:threats}).
Then, we analyze limits on user privacy by features reported by \toolname (Section~\ref{sec:uniqueness}).
To do that, in Section~\ref{sec:real-world-privacy} we focus on the larger, $M$ dataset of 810 users and 80k ads, used in~\cite{imcRTB2017} with the same features as our system's.
This dataset allows us to study real users' distributions of features, and how user anonymity changes with feature aggregation.
Finally, in Section~\ref{sec:tradeoff}, we study the inherent trade-off between feature aggregation and price estimation accuracy.

\subsection{User de-anonymization threats}\label{sec:threats}
Below, we discuss possible de-anonymization threats that the system could face during data reporting, and how the system design prevents them from materializing.

\point{Threat 1: De-anonymization using users' online behavior or demographics.}
Most users have a consistent online behavior, meaning they tend to visit the same websites through their week (e.g., a user may visit {\tt cnn.com} to read daily news, vs. another user who does that via {\tt boston.com}).
This behavior makes the user identifiable, as already shown in~\cite{wit}.
In addition, the 1st-party domain that an ad was detected on, could be a private or sensitive domain or a domain that other users are very unlikely to visit.
Also, browsing behavior could be linked to particular demographics like age and gender.
We address these threats by not reporting the 1st-party domains visited by users but their corresponding IAB categories~\cite{iab_lab, iab_taxonomy}.
In this way, multiple different sites can appear under the same IAB (\eg all news sites will be under the same IAB).
Furthermore, the users may choose to provide a self-defined gender (binary) and age (aggregated in coarse-grained 10-year bins).

\point{Threat 2: De-anonymization using advertising-related activity.}
Users could be monitored by different trackers while browsing the web~\cite{englehardtonline, 10.1145/3442381.3450056, bashir2018diffusion}, with user-specific IDs and other identifiers associated with them.
In order to reduce the probability of de-anonymization of a user, we do not report any identifiers or trackers, or even advertisers involved.
Instead, we extract metadata from the nURL such as the winner DSP (i.e., the intermediary entity between advertiser and ADX), the price keyword of the ADX, the price value and ad size, if available.
We also report if cookie syncing was detected.

\point{Threat 3: De-anonymization using grouped records into user sessions.}
An attacker may attempt to de-anonymize a user, by assuming that consecutive records of incoming data could be reported by the same user.
The system proceeds to the following actions to reduce or even eliminate such a threat:
(1) As explained in Section~\ref{sec:features}, the reporting client does not send data for every detected ad, but instead collects a set of records which it then shuffles and sends to the server at random times.
This strategy also breaks possible time linkage between records, as they are not reported as soon as they are created.
(2) The server randomly shuffles incoming data with stored data frequently, depending on the amount of data stored.
This enables the server to break any possible sessions or relation that records have.

\noindent \textbf{Finding:} The privacy-preserving design of \toolname protects users from de-anonymization attacks in various typical or extreme scenarios.

\subsection{User uniqueness via surprisal analysis}\label{sec:uniqueness}

Overall, the system design can protect users' anonymity by removing identifiers or attributes that contain unique examples of users' activity, reducing the granularity of reported data (with respect to frequency of reporting or detail), and randomizing data reporting and storing to avoid possibility of linking records together into sessions.
However, it is important to study the bounds of anonymity that can be achieved, when features reported to the server are considered together while attempting to de-anonymize a user.

In this section, we formally investigate this problem.
In particular, we are interested to study what granularity of reporting is low enough to protect the anonymity of each user, while still maintaining full or partial utility of the reported data.
For this investigation, we employ two commonly known tools to understand this trade-off.
First, we study how unique a user is, given a particular reporting setup (\ie combination of features and granularity classes).
and (similar to other works~\cite{Eckersley:2010:UYW:1881151.1881152}) we estimate this by measuring the number of \emph{surprisal bits} that such a setup achieves.
Second, we compute the number of users who are expected to be found or collide in a given reporting setup.
The second study is a classic k-anonymity analysis, in which we measure how many users have similar behavior and are likely to report the same combination of features and granularity classes.
Next, we offer these analyses of surprisal and k-anonymity for the features considered in Section~\ref{sec:features}.

\point{Background on Information Theory:}
In information theory, given an event $E$ with probability $P(E)$, surprisal is 
$I(E) = -\log_2{(P(E))}$,
and is measured in bits; the higher the surprisal the more unique an event is.
Something that is certain has no surprisal (0 bits); flipping a fair coin is associated with 1 bit of surprisal; winning the lottery has 24 bits of surprisal.
In our case, higher surprisal of a user's reporting data means they can be uniquely identified easier.

If the said event E is dependent on $N$ independent attributes or features ($i \in N$), then the overall probability of $P(E)$ can be expressed as:
\begin{equation}\label{eq:uniform}
P(E) = p_1 \times p_2 \times \dots p_i \dots \times p_N.
\end{equation}
If we assume that each feature $i$ has different discrete classes, and they are equally likely to appear (i.e., they are governed by a uniform distribution), then every feature $i$ has a probability 
$p_i=1/{\#}\ {of}\ {distinct}\ {classes}\ {of}\ {feature}\ i$.

\point{Uniqueness using Uniform Distributions:}
Under the assumption of a uniform distribution of classes of features (Eq.~\ref{eq:uniform}), all the combinations of features have the same probability to occur.
This assumption allows us to study the problem of data reporting and define baselines of features and their classes to be reported.

In a more realistic scenario, an event $E$, or in our case experimental setup $E$ (that dictates specific class for each of the features), is dependent on $N$ independent features that are governed by real distributions.
Thus, the above probabilities are not equal for all classes of a given feature.
Instead, for class $j$ of feature $i$ we have 
$p_{ij}=PDF_i(E_{ij})$,
where $PDF_i()$ is the real (or observed) probability distribution function of feature $i$, and $PDF_i(E_{ij})$ is the probability returned by this function for the event $E$ to occur at class $j$ of this feature $i$, $E_{ij}$ (\ie the particular level or class that the experimental setup takes ($j$) for the feature $i$).
For example, feature $i$ could be the time of day, and the different classes of this feature could be the 24 hours available ($j$).
Then, the probability of an ad to be detected at the 11am slot, is given by the $PDF_{i=\textrm{time of day}}(E_{ij=11am})$.

We can compute overall $P(E)$ for a specific combination of features $i \in N$ and classes $j$:
\begin{equation}\label{eq:real}
P(E)=PDF_1(E_{1j})\times\dots\times PDF_N(E_{Nj})=\prod_{i=1}^{N} PDF_i(E_{ij})
\end{equation}
where:																	$E$ = the experimental setup of interest, comprised of a selected class $j$, for each feature $E_i$,
$PDF_i(E_{ij})$ = probability of $E$ to occur, as computed from the PDF of feature $i$ for its class $j$,		
$P(E)$ = overall probability of $E$ to occur.

\point{User uniqueness via surprisal analysis assuming uniform distributions:}
Following the assumption of a uniform distribution of classes of features (Eq.~\ref{eq:uniform}), all the combinations of features have the same probability to occur in our dataset.
This assumption allows us to study the problem of data reporting and define some baselines of features and their classes that we could report.
In Table~\ref{table:surprisal-uniform} we compute the theoretical surprisal bits of a system that has the features mentioned earlier and are shown in individual rows in the table.
The granularity of reporting, \ie the number of independent and uniformly distributed classes is reported in each column for each row.

We begin by assuming that the \toolname reports back the features exactly as collected without any generalization.
This scenario is hypothetical, but can give us a baseline understanding of what we are dealing with respect to surprisal bits.
We assume that the age ranges from 0 to 99 years, the user could be from any of the 240 different countries of the world and that the reporting is in 1-hour bins.
We also assume that the client reports back the 1st-party domains they browse through (\eg 500 distinct values) and the trackers they have encountered and performed cookie synchronization (\eg 200 distinct values).
The price keyword can take only 1 distinct value because each DSP is using one keyword, but the price value can obviously vary.
Thus, we assume this price value can take 200 distinct values.
In this hypothetical scenario, there is no aggregation so the surprisal, as expected, is high ($>60$ bits).

\begin{table}[t]
	\centering
	\caption{Number of surprisal bits for different distinct classes of each feature, assuming uniform distribution of appearance of each class, for each feature.}
	{\footnotesize
		\begin{tabular}{lrrrrrrrrr}
			\toprule
			\textbf{Feature} & \multicolumn{9}{c}{\textbf{Number of classes}} \\
			\midrule
			Gender & 2 & 2 & 2 & 2 & 2 & 2 & 2 & 2 & 2 \\
			Age & 100 & 5 & 5 & 5 & 5 & 5 & 5 & 5 & 5 \\
			Country & 240 & 240 & 240 & 240 & 240 & 120 & 120 & 25 & 25 \\
			Time Of Day & 24 & 24 & 8 & 8 & 8 & 8 & 8 & 8 & 8 \\
			Day Of Week & 7 & 7 & 7 & 7 & 7 & 7 & 7 & 7 & 7 \\
			Cookie Sync & 200 & 200 & 200 & 2 & 2 & 2 & 2 & 2 & 2 \\
			DoNotTrack & 2 & 2 & 2 & 2 & 2 & 2 & 2 & 2 & 2 \\
			adFormat & 20 & 20 & 20 & 20 & 20 & 20 & 20 & 20 & 3 \\
			winnerDSP & 200 & 200 & 200 & 200 & 200 & 200 & 200 & 200 & 200 \\
			category & 500 & 500 & 500 & 500 & 30 & 30 & 30 & 30 & 30 \\
			Price Keyword & 1 & 1 & 1 & 1 & 1 & 1 & 1 & 1 & 1 \\
			Price Value & 200 & 200 & 200 & 200 & 200 & 200 & 50 & 50 & 3 \\
			Surprisal Bits & 60.2 & 55.8 & 54.3 & 47.6 & 43.6 & 42.6 & 40.6 & 38.3 & 31.5 \\
			\bottomrule
		\end{tabular}
	}
	\label{table:surprisal-uniform}
\end{table}

Next, we assume that the tool performs aggregation of the distinct classes of some of the features, to explore a more realistic setup.
For example, by just grouping the age into 5 distinct ranges, we reduce the surprisal bits by 5.
Grouping the reported time of day in 3-hour intervals seems to have only a small impact to the overall surprisal rate.
In contrast, by eliminating the cookie synchronization DSP, and reporting back only if CS took place (2 distinct values) we greatly reduce the overall surprisal by almost 7 bits.
Also, if instead of reporting the 1st-party domain, we report only 30 distinct IAB categories, we reduce the surprisal by over 4 bits.

In the far right case, in which all possible features are aggregated, we can effectively halve the initial surprisal rate.
But this can impact greatly the utility of our data.
In this extreme case, no actual value is sent.
Instead, the client only reports if the price was low, medium or high.
This is the same case with the ad format, where the client reports if the ad was small, medium or large.
Finally, instead of the actual country, the client reports only a country zone (\eg Central EU, North America, etc.).
The combination of distinct features and their impact on the surprisal rate can be seen in Table~\ref{table:surprisal-uniform}.
From the previous threat analysis and the computation of surprisal bits assuming uniform distribution of classes, we can extract lessons on what features can be reported and with how many classes each.

\subsection{Real-world data privacy analysis}\label{sec:real-world-privacy}

Building on the previous theoretical analysis, we conduct an anonymity analysis of the reported features on a real user dataset.
However, in this case, we employ the formula of Eq.~\ref{eq:real}, where the probability of a class (or level) in a feature is no longer uniform, but is computed using the probability distribution function of the feature.
The dataset used is $M$, with $80k$ ad impressions from 810 different users, each entry containing 27 different features.
Out of these 27, we selected the ones reported back from \toolname as well, ending up with a set of 8 features: user location, day of week, time of day, ad size, ADX involved, IAB category of 1st-party, price keyword and price value.
Based on the classes in each feature, we calculate the probability of each class.

We start our analysis by examining a pessimistic scenario, which assumes an experimental setup where all features are reported at their classes with lowest probability observed.
That is, we take the lowest probability class for each feature and compute the surprisal bits of this scenario.
Indeed, the surprisal bits for a user who reports such unlikely event are about $60$ bits.
Interestingly, this number is very close to our previous theoretical worst case scenario, where no kind of anonymization was applied.

\begin{figure}[t]
	\centering
	\includegraphics[width=0.7\linewidth]{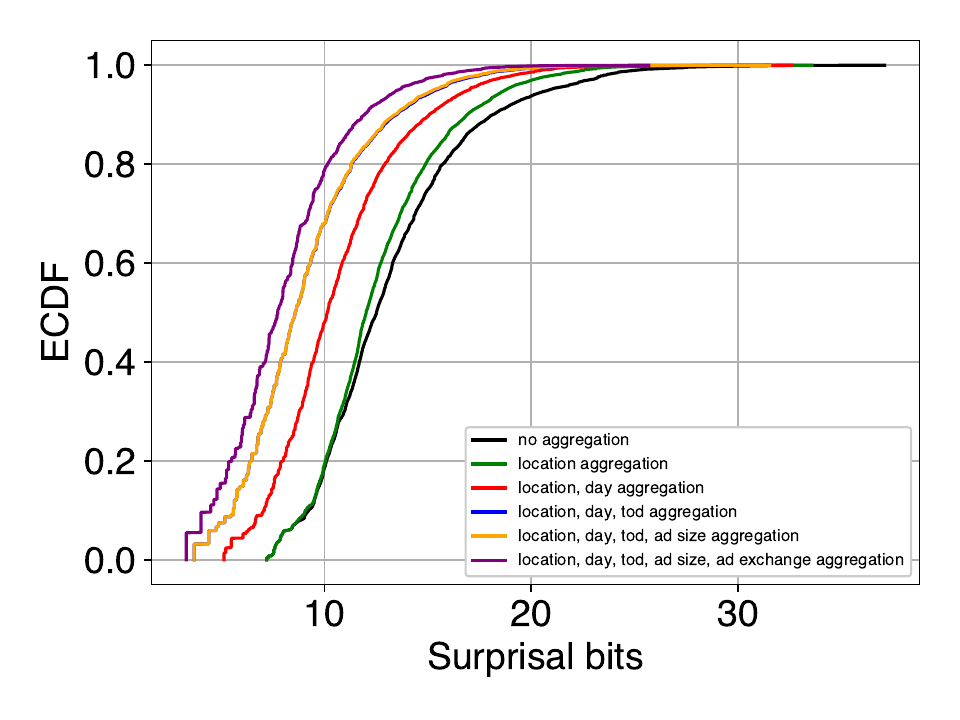}
	\caption{Surprisal bits after incremental feature aggregation.}
	\label{fig:surpr_aggr_real}
\end{figure}

Next, we proceed to analyze the scenarios that did happen in the real dataset, and compute their surprisal bits.
In Figure~\ref{fig:surpr_aggr_real} (black line), we show the CDF of the surprisal bits for the observed setups.
All setups have fewer than $37$ bits, and $95\%$ of the cases have fewer than $21$ bits.
Also, the median setup has $12.5$ bits.
Indeed, someone could argue that the high surprisal bit rate for a small portion of entries could expose some users.
We remind the reader that in the absence of any sort of PII, the link to specific users can be difficult if not impossible inside this database.

Similarly with the analysis for uniform distribution, we can attempt to improve anonymity of users by performing aggregation of features to fewer classes.
We demonstrate this effort in Table~\ref{table:aggr_features_real}, where we apply incremental aggregation of each feature into larger groups or levels.
For example, locations can be reported in higher abstraction such as districts instead of cities, days can be reported as weekdays vs. weekends, 4-hour slots can be grouped into larger bins, etc.
In Figure~\ref{fig:surpr_aggr_real}, we show the CDF of the overall surprisal bits for the different setups included in the real dataset, and different aggregation efforts of Table~\ref{table:aggr_features_real}.
Interestingly, when we perform aggregation for the different features, we see a steady reduction of surprisal bits across all cases, besides the location aggregation which is effective for a portion of the ads.
This is due to the fact that in such ads, the locations are popular and summarizing them does not have an impact in the surprisal rate.
After applying aggregations in all classes, the surprisal rate of the median setup drops from $12.5$ bits to $7.7$ bits, and the $95\%$ case drops from $21$ bits to $13.7$ bits.

\noindent \textbf{Finding:} With feature aggregation, the median surprisal bits under various distributions of classes (uniform or real) can be halved, in comparison to no aggregation scenarios.
Also, location aggregation may not reduce user uniqueness as much as other features such as aggregation by time of day, or by day of week.

We continue this investigation by performing a k-anonymity~\cite{kanonymity_sweeney,papadopoulos2013k} analysis of this real dataset, to understand at what level it is satisfied, based on the different classes and aggregation performed.
K-anonymity is used as privacy criterion in real applications such as the ``Family Educational Rights and Privacy Act'' (FERPA) of USA~\cite{ferpa}, and the ``Guidelines for De-identification of Personal Data'' of South Korea~\cite{koreanprivacy}.
We consider a realistic scenario where the attacker has limited access per ad detected, without any knowledge of the user's browsing behavior (\ie does not have IAB categories reported).
In this scenario, the attacker groups the entries found in the dataset based on the features of location, day of the week and time of the day (i.e., does not have access to other features).
In this attack scenario and the different setups tested, a minimum k-anonymity of k=6 users is satisfied, and a median entry (\ie ad) can be mapped to $k$=35 users.
In general, these scores are within the range of $k$=3--10 reported in~\cite{healthdatareporting2017bcm} and applied for electronic health records, lending support to the applicability in our scenarios as well.
In future versions of the tool, we will also consider other techniques such as Differential Privacy~\cite{dwork2006dp}, that can provide more tight privacy guarantees to the end user.

\noindent \textbf{Finding:} With feature aggregation, a median (30-45)-anonymity can be achieved.

\begin{table}[t]
	\caption{Feature class aggregation vs. surprisal bits (assuming uniform distribution of classes), and vs. performance of ML modeling.}\vspace{-0.2cm}
	{\footnotesize
		\begin{tabular}{lrrrrrr}
			\toprule
			\textbf{Feature} & \multicolumn{6}{c}{\textbf{Number of classes per feature}} \\ 
			\midrule
			user location	& 184 & 26  & 26  & 26  & 26  & 26 \\
			day of week	& 7   & 7   & 2   & 2   & 2   & 2  \\
			time of day	& 6   & 6   & 6   & 2   & 2   & 2  \\
			ad size		& 17  & 17  & 17  & 17  & 3   & 3  \\
			ad exchange	& 149 & 149 & 149 & 149 & 149 & 15 \\
			cookie syncing	& 2   & 2   & 2   & 2   & 2   & 2  \\
			IAB category	& 25  & 25  & 25  & 25  & 25  & 25	\\ 
			\midrule
			Surprisal Bits	& 29.9	& 27.0	& 25.2	& 23.7	& 21.2	& 17.8	\\ \midrule
			\multicolumn{7}{c}{\textbf{ML performance vs. feature aggregation}} 	\\ \hline
			AUC of ROC	& 0.858	& 0.854	& 0.820	& 0.798	& 0.798	& 0.786	\\
			\bottomrule
		\end{tabular}
	}
	\label{table:aggr_features_real}
\end{table}

\subsection{User Anonymity vs. Price Modeling }\label{sec:tradeoff}
We close this investigation by looking into the inherent trade-off between aggregation of feature classes, and modeling of prices at the server using machine learning (ML) methods.
We use a decision tree model and 4 equidistant classes for the RTB prices detected and measure AUCROC, a standard ML performance metric.
We use this model with features aggregated at different levels, as examined earlier.
The results are shown at the bottom of Table~\ref{table:aggr_features_real}.
We find that the AUC is not greatly impacted by the aggregation performed (8.4\% decrease) when comparing the no-aggregation scenario (1st column) vs. the fully aggregated scenario (last column).
These results show that \toolname can do intense feature aggregation at the client before reporting to the back-end, without significantly affecting the server's ML model performance.
Note here, that another formal way to achieve anonymization of user data with provable guarantees is by applying techniques such as Differential Privacy (DP)~\cite{dwork2006dp}.
In such method, Laplacian or Gaussian noise can be added on each feature to be reported, so that the probability of existence of individual rows into the dataset is obfuscated up to a certain, and system-controlled degree.
However, the problem with such technique, in relation to our setup, is that the full distribution of each feature must be known, before proper noise level is selected for each feature to be protected.
For the reconstruction at the server of the full distribution in an anonymous fashion, complex and costly methods such as the one used in EyeWnder~\cite{iwnder} can be employed.
Alternatively, emerging, privacy-preserving ML methods such as federated learning coupled with differential privacy~\cite{geyer2017dp-fl} can be used, to allow the server the building of a global ML model across users, without the need to report raw or aggregated user data, but only pre-models with added DP-noise.
We will investigate such techniques in the future.

\noindent
\textbf{Finding:} The \toolname client can perform high feature aggregation before reporting data with a minimal impact on the performance of the ML price model used in the system.

\section{Related Work}

In~\cite{imcRTB2017}, authors explore the cost advertisers pay to deliver an ad to the user in  RTB auctions and how the personal data can affect this cost.
The authors proposed a methodology to compute the total cost paid for the user even when advertisers hide the charged prices. Finally, they evaluated their methodology by using data from a large number of volunteering users. 
They find that advertisers paid a total of around 25 CPM to deliver ads to the average user across a year. 
In~\cite{papadopoulos2018cost},  authors use the same methodology to measure the costs of digital advertising on both the user's and the advertiser's side in an attempt to compare how fairly these costs are distributed between the two. In particular, they compare the cost advertisers pay in RTB with the costs imposed on the dataplan, the battery efficiency and the privacy of the specific user.
In~\cite{lukasz2014selling-privacy-auction}, authors detected RTB notification URLs and extracted the value of the auctioned ad.
They made an extensive study on the RTB ecosystem and estimated the value of user's private data based on the cleartext price notification URLs. They found that the average price of an ad is 0.0001\$-0.004\$, depending on the user and ad campaign.  

In~\cite{followTheMoney}, authors use a dataset of users' HTTP traces and provide rough estimates of the relative value of users by leveraging the suggested bid amounts for the visited websites, based on categories provided by the Google AdWords.
FDTV~\cite{gonzalez2017fdvt} is a tool to inform users in real-time about the monetary value of the personal information associated to their Facebook activity.
Contrary to our work, FDVT is obtaining prices from the Facebook AdPlanner and thus is relevant only for Facebook advertising. 
In \cite{CSync}, authors used a heuristic-based mechanism to detect information exchanged between advertisers through CS. They concluded that 97\% of the users are exposed to CS at least once and ad-related entities participate in more than 75\% of the overall synchronizations. In~\cite{Papadopoulos:2018:ECM:3193111.3193117} authors demonstrated how  this technique may leak user's cookie IDs and browsing history to a snooping ISP even when user uses TLS and secure VPN services. 

Bashir \etal in~\cite{bashir2018diffusion}, studied the diffusion of user tracking caused by RTB-based programmatic ad-auctions.
Results of their study show that under specific assumptions, no less than~52 tracking companies can observe at least~91\% of an average user's browsing history.
In~\cite{bashirtracing}, the same group tried to enhance the transparency in ad ecosystem with regards to information sharing, by developing a content agnostic methodology to detect client- and server- side flows of information between ad exchanges and leveraging retargeted ads.
By using crawled data, they collected 35.4k ad impressions and identified 4 types of information sharing between ADXs. 

In~\cite{acquisti2013privacy}, authors discuss the value of privacy after defining two concepts (i)~\emph{Willingness To Pay}: the monetary amount users 
are willing to pay  to  protect  their  privacy, and (ii)~\emph{Willingness  To  Accept}: the compensation that users are willing to accept for their privacy loss.
Two user-studies~\cite{bigmac,staiano2014moneywalks} have measured how much users value their own offline and online personal data, and consequently how much they would sell them to advertisers.
In~\cite{forSale}, they propose ``transactional'' privacy to allow users to decide what personal information can be released and receive compensation from selling them. 
In \cite{header_bidding}, authors measure the prevalence, pricing dynamics and characteristics of a different type of programmatic ad-buying namely header bidding.
In \cite{hb_userStudy} the authors conduct an analysis of the header bidding advertising ecosystem using a set of real users. Header Bidding is a relatively newly introduced ad-buying protocol which gains popularity among advertisers.
In our present study, we focus on the widely used RTB protocol namely \emph{waterfalling}, since it is more mature and commonly accepted among all publishers.
In~\cite{adconcerns3}, authors measured how much value online publishers derive from behavioral ad targeting.
Their results show that publishers make just 4\% more revenue when including trackers in their websites.

\section{Ethical Considerations}
\label{sec:ethics}
The execution of this work has followed the principles and guidelines of how to perform ethical information research and the use of shared measurement data~\cite{dittrich2012menloreport,rivers2014ethicalresearchstandards}.
In particular, this study paid attention to the following dimensions.

\noindent \textbf{Respect for Persons:}
We received \textit{informed consent} from users, who were provided with information about our study and tool by visiting our page, and before downloading and installing our plugin.
These users were \textit{volunteers} and expressed their \textit{comprehension} to the requirements for participating in the study and allow us to collect their anonymized data.
We treated all volunteers \textit{equally and with justice}, and without prejudice to any gender, age or other demographic characteristic.

\noindent \textbf{Respect for Law and Public Interest:}
In accordance to the GDPR and ePrivacy regulations, and by designing our tool to be privacy-by-design, we do not collect any kind of identifiers that can de-anonymize users of the tool.
Also, we do not share with any other entity any data of users collected by our tool, and especially any sensitive personal data that may have inadvertently collected without our knowledge.
Furthermore, user data have been stored in secured manner, according to GDPR policies.

\noindent \textbf{Beneficence:}
During the design of the experimental tool and the decision on what data to be collected, we assessed the \textit{benefits to society} from the findings of our study and made significant efforts to reduce, or at least \textit{mitigate any harms} that could be experienced by any individual participant.
In fact, and as explained in our system design section, we made significant efforts to design a tool that does not impact the users' experience while browsing the Web, nor has adverse effects to the ad-ecosystem while intercepting the notification URLs sent by ad-exchanges to DSPs for charging and information purposes.

\section{Summary \& Conclusion}
\label{sec:discussion}

In this paper, we presented the design, implementation and deployment  of \toolname, a first-of-its kind, full fledged system that allows any user to learn in real time while browsing how much money the RTB ecosystem pays to show them ads.
We operated the system for a year, and it was used by \rusers users.
During this time period, we collected RTB-price metadata, and using past reported numbers for RTB ad-prices in mobile and desktop, in this paper, we make the following findings regarding RTB prices and their evolution over time:

\begin{itemize}[itemsep=0pt, leftmargin=\parindent]
	\item Desktop RTB prices have increased 4.6$\times$ within 7 years.
	\item During weekends and Mondays, the ad-prices tend to be higher than weekdays.
	\item Ad-prices are higher in the morning hours than afternoon or evening hours.
	\item The most popular publisher categories (using IAB) are News, and Technology/Computing.
	\item The most expensive publisher categories (using IAB) are Sports, and Food/Drink.
	\item Advertisers pay 2.4$\times$ more to target women than men.
	\item Younger users (25-34 years old) cost 2.5$\times$ more to be reached than older groups (35-44 years old).
	\item USA, Switzerland, and Spain have higher median ad prices across the 11 countries in our dataset.
	\item Performing cookie syncing before an RTB auction leads to increased median prices by 19\%.
\end{itemize}

In addition, we performed a privacy evaluation of the system, to identify possible threats against user's anonymity.
We measured the limits of user anonymity with a uniqueness study via surprisal and k-anonymity analysis.
Finally, we studied the trade-off of anonymity via feature aggregation vs. performance of price modeling with ML methods.
In summary, the main takeaways from the privacy evaluation of this tool were:

\begin{itemize}[itemsep=0pt, leftmargin=\parindent]
\item \toolname's privacy-preserving design protects users from typical and extreme de-anonymization attacks.
\item With feature aggregation, the median surprisal bits under various distributions of classes (uniform or real) can be halved to 7.7bits, in comparison to no-aggregation scenarios.
\item Location aggregation does not reduce user uniqueness as much as other features (\eg time of day or day of week).
\item Feature aggregation enables k-anonymity with median $k$=30-45.
\item \toolname's client can do high feature aggregation before reporting with minimal impact on the ML price model.
\end{itemize}

We have demonstrated that the \toolname tool can successfully capture the state of RTB across the market, by verifying our findings (historical or not) with industry and other technical reports.
Interestingly, \toolname does this without jeopardizing the end-user's privacy, but instead offers a looking glass on RTB via cheap and distributed experimentation.
We envision that in the future, the tool will be further used by many end-users, privacy researchers and auditors, who can take advantage of its simple functionalities to increase transparency in the RTB ad-ecosystem and its obscure practices of user modeling and ad-costs.

\section*{Acknowledgements}
This project received funding from the EU H2020 Research and Innovation programme under grant agreements No 830929 (CyberSec4Europe), No 830927 (Concordia), No 871793 (Accordion), No 871370 (Pimcity) and Marie Sklodowska-Curie grant agreement No 690972 (Protasis).
These results reflect only the authors' view and the Commission is not responsible for any use that may be made of the information it contains.

\bibliographystyle{unsrt}
\bibliography{main}

\received{August 2021} 
\received[revised]{October 2021}
\received[accepted]{November 2021}

\end{document}